\documentclass[11pt]{article}
\pdfoutput=1

\usepackage[letterpaper,
width=6.5in,
left=1in,
top=1.3in,
bottom=1.15in,
headheight=19pt,
]{geometry}

\usepackage{url}            
\usepackage{graphicx}
\usepackage{wrapfig}
\usepackage{amssymb}
\usepackage{amsmath}
\usepackage[english]{babel}
\usepackage{tabularx} 

\usepackage{enumitem}
\usepackage[normalem]{ulem}
\usepackage{listings}
\usepackage{./protobuf/lang}  
\usepackage{./protobuf/style} 
\usepackage{wrapfig}
\allowdisplaybreaks
\usepackage[hang,flushmargin,bottom]{footmisc}
\usepackage{mathtools}
\DeclarePairedDelimiter\ceil{\lceil}{\rceil}
\DeclarePairedDelimiter\floor{\lfloor}{\rfloor}
\usepackage[ruled, algosection, noend, linesnumbered]{algorithm2e}
\SetAlFnt{\small}
\SetAlCapFnt{\large}
\SetAlCapNameFnt{\large}

\SetCommentSty{mycommfont}

\usepackage{caption}
\DeclareCaptionFont{xipt}{\fontfamily{fvs}\fontsize{9}{11}\selectfont}
\usepackage[labelfont={xipt,bf},textfont={xipt},font={xipt}]{caption}

\newtheorem{theorem}{Theorem}[section]
\newtheorem{corollary}{Corollary}[theorem]
\newtheorem{lemma}[theorem]{Lemma}

\usepackage{multirow}
\usepackage[square,sort,comma,numbers]{natbib}

\definecolor{darkgreen}{RGB}{0,192,0}

\graphicspath{{./}{./figures2/}}

\newcommand\hmm[1]{\ifnum\ifhmode\spacefactor\else2000\fi>1000 \uppercase{#1}\else#1\fi}
\newcommand{\Architecture}{Flow\xspace}
\newcommand{\entropy}{\hmm{s}ource of randomness\xspace}
\newcommand{\techterm}[1]{{\sffamily\selectfont{#1}}}

\usepackage{hyperref}       
\hypersetup{
	colorlinks=true,       
	linkcolor=[rgb]{0,0.4,0.5},          
	citecolor=[rgb]{0,0.3,0.3},        
	urlcolor=[rgb]{0,0.3,0}           
}

\newcommand{\collector}{N_{c}}
\newcommand{\collectorsNum}{n_{c}}

\newcommand{\consensusNum}{n_{s}}
\newcommand{\cluster}{\texttt{Cls}}
\newcommand{\clusterSize}{k}
\newcommand{\randomBeacon}{r}
\newcommand{\clusterNum}{c}
\newcommand{\nodesNum}{n\xspace}

\newcommand{\consThr}{\frac{2}{3}}
\newcommand{\consensus}{HotStuff\xspace}

\newtheorem{definition}{Definition}[section]

\newcounter{protocol}
\newenvironment{protocol}[1]
  {
   \noindent
   \tabularx{\linewidth}{@{} X @{}}
    \hline
    \refstepcounter{protocol}\textbf{Protocol \theprotocol} #1 \\
    \hline}
  { \\
    \hline
   \endtabularx
   \par\addvspace{\topsep}}

\newcommand{\sbline}{\\[0\normalbaselineskip]}

\title{\Architecture: Separating Consensus and Compute\\[5pt]\Large
	-- Block Formation and Execution --}
\date{} 

\begin{document}
\maketitle

\begin{center}
    \vspace{-20pt}
	\begin{tabular}{c@{\hspace{60pt}}c}
		\multicolumn{2}{c}{Dr.\ Alexander Hentschel} \\
	    \multicolumn{2}{c}{\footnotesize\texttt{alex.hentschel@dapperlabs.com}} \\[10pt]
		Dr.\ Yahya Hassanzadeh-Nazarabadi & Ramtin Seraj \\
	    \footnotesize\texttt{yahya@dapperlabs.com} & \footnotesize\texttt{ramtin.seraj@dapperlabs.com} \\[10pt]
		Dieter Shirley & Layne Lafrance \\
		\footnotesize\texttt{dete@dapperlabs.com} & \footnotesize\texttt{layne@dapperlabs.com}
	\end{tabular}
	\\[10pt]
\end{center}
\vspace{30pt}

\begin{abstract}
\noindent
Most current blockchains are built as a homogeneous system, comprised of \textit{full nodes}, which are responsible for all tasks: collecting the transactions, block formation, consensus, and transaction execution.
Requiring all full nodes to execute all tasks limits the throughput of existing blockchains, which are well documented and among the most
significant hurdles for the widespread adoption of decentralized technology \cite{Kyle2016, BBC:CK_slows_Ethereum:2017, Richard_et_al:Bitcoin_Ethereum_Scalability:2019}.

This paper is a follow-up on our previous proof-of-concept, \Architecture \cite{shirley2019flow}, a pipelined blockchain architecture, which separates the process of consensus on the transaction order from transaction computation. As we experimentally showed in our previous white paper, this provides a significant throughput improvement while preserving the security of the system. \Architecture exploits the heterogeneity offered by the nodes, in terms of bandwidth, storage, and computational capacity, and defines the roles for the nodes based on their tasks in the pipeline, i.e., Collector, Consensus, Execution, and Verification. While transaction collection from the user agents is completed through the bandwidth-optimized Collector Nodes, the execution of them is done by the compute-optimized Execution Nodes. Checking the execution result is then distributed among a more extensive set of Verification Nodes,  which confirm the result is correct in a distributed and parallel manner. In contrast to more traditional blockchain  architectures, \Architecture's Consensus Nodes do not execute the transaction. Instead, Verification Nodes report observed faulty executions to the Consensus Nodes, which adjudicate the received challenges and slash malicious actors. 

In this paper, we detail the lifecycle of the transactions from the submission to the system until they are getting executed. The paper covers the Collector, Consensus, and Execution role. We provide a protocol specification of collecting the transactions, forming a block, and executing the resulting block. Moreover, we elaborate on the safety and liveness of the system concerning these processes.

\end{abstract}

\newpage
\tableofcontents

\newpage
\section{Introduction}
\label{flow_wp2:section_introduction}

\subsection{Architecture Overview}

In most conventional blockchains, all the operational tasks are executed by \textit{full nodes}, i.e., each full node is expected to collect and choose the transactions to be included in the next block, execute the block, come to a consensus over the output of the block with other full nodes, and finally sign the block, and append it to the chain. The structure resembles the single-cycle processing architecture of the early computers, where one instruction (i.e., block) is executed per step. On the other hand, the pipelined structure offered by \Architecture resembles the pipelined architecture of modern CPU design. While some unit of \Architecture is executing the current block, the pipeline architecture allows another unit to generate the next block, which results in an improvement over the throughput.   

In contrast to the majority of existing blockchains, which operate on the assumption that nodes are homogeneous, \Architecture considers resources heterogeneous in their network bandwidth, computation power, and storage capabilities. The \Architecture blockchain is built on a diverse set of heterogeneous nodes, each performing a specific set of tasks associated with their role(s). Specifically, there are four different roles: \textit{Collector Nodes, Consensus Nodes, Execution Nodes, and Verification Nodes},  which handle the collection of transactions, generating blocks, executing blocks, and verifying the execution results, respectively. \Architecture allows a high level of participation in Consensus and Verification by requiring only a high-end consumer internet connection while leveraging large-scale data centers to operate as Execution Nodes. All types of nodes in \Architecture receive compensation through crypto-economic incentives. Prior research  indicates that exploiting the heterogeneity of the nodes and taking the pipelined approach results in a throughput improvement by a multiplicative factor of $56$ \cite{shirley2019flow} compared to homogeneous architectures using only full nodes. We also showed that the decentralization and safety of the network is preserved by the \Architecture architecture \cite{hentschel2019flow}.

It is worth noting that the separation of consensus from execution is done based on the separation of the \emph{subjective} from \emph{objective} operations. The \textit{Subjective} operations are those that do not have a deterministic result. Instead, they require a \textit{consensus} to be reached over the result. An example of a \textit{Subjective} act is the set of transactions to be included in the next block. On the other hand, the \textit{Objective} operations are the ones with a deterministic result and do not require the consensus over their result. An example of an \textit{Objective} act is the execution of a transaction that transfers some tokens between two accounts. In \Architecture, the separation of consensus and execution is done by assigning the \textit{Subjective} tasks to the Consensus Nodes, and the \textit{Objective} tasks to the Execution Nodes. Hence, even a few data centers may run the Execution Nodes. Nevertheless, their computation results are deterministic, verifiable, attributable, punishable, and recoverable, and do not compromise the decentralization of the system. We iterate over these terminologies in the rest of this section.

\subsection{Design Principles}
For \Architecture to be secure against Byzantine failures and attacks, we identify the following core architectural principles. In this paper, by an honest actor, we mean a node that exactly follows protocol associated with its role and never deviates from the protocol. 
\begin{itemize}
    \item \textit{Detectable}: 
    Any honest actor in the network can detect deterministic faults and prove the fault to all other honest nodes. The proof only requires asking other honest nodes to redo the faulty parts. 
    \item \textit{Attributable}: 
    For performance reasons, many tasks in \Architecture are assigned randomly to a group of nodes. While the assignments are random for security reasons, it is based on \textit{Verifiable Random Functions (VRFs)} \cite{Micali:1999:VRFs}. Hence, any fault upon detection is also attributable to the responsible node(s) of the associated task.
    \item \textit{Punishable}:
    Every node participating in the \Architecture network must put up a stake, which is slashed when a detected fault is attributed to the node.
    Reliably punishing errors via slashing is possible because all errors in deterministic processes are detectable\footnote{%
        In \Architecture, nodes check protocol-compliant behavior of other nodes by re-executing their work.
        In most cases, verification is computationally cheap, with the noticeable exception of computing all transactions in a block. 
        We describe in our previous white paper the procedure of distributing and parallelizing the verification of the computation of the blocks \cite{hentschel2019flow}. Hence, each Verification Node only performs a small fraction of the overall block computation. 
    }
    and attributable.
    \item \textit{Recoverable}:
    The \Architecture protocol contains specific elements for result verification and resolution of potential challenges.
    These elements act as countermeasures against the \textit{attempts} of the malicious nodes to inject faults into the system. The probability of successfully injecting such faults is negligible.
\end{itemize}

\subsubsection{Safety and Liveness}
As a distributed multi-process system, the correctness of \Architecture protocol sets are evaluated against their \textit{safety} and \textit{liveness} \cite{lamport1977proving}. A distributed multi-process system is safe against a set of features, if it provides guarantees that those features never happen in the system, unless with a negligible probability in some specific system parameter. Similarly, the system shows liveness respect to a set of features, if it provides guarantees that those features always persist, unless with a negligible probability in some specific system's parameter. In this paper, we formally identify the liveness and safety features of the introduced protocols. We prove \Architecture is safe and live in the presence of a limited fraction of Byzantine actors in each role.

In the design of \Architecture, we \textit{prioritize safety over liveness} in case of a network split. A network split happens when at least a non-empty subset of nodes in the system is not accessible by rest of the nodes. In the extremely unlikely circumstance of a large-scale network split, where no connected subset of the network includes enough nodes to make forward progress, we allow the network to halt. This preserves the safety of the network even in the face of intentional network disruption, such as with an Eclipse attack \cite{heilman2015eclipse}. 

\subsection{Assumptions}
The \Architecture architecture makes the following set of assumptions:
\begin{itemize}
    \item \textit{Consensus and Authentication} 
    \begin{itemize}
        \item All nodes participating in the system are known to each other. 
        \item Each node is authenticated through its unforgeable digital signature.
        \item To reduce the computational cost and preserve the sustainability of the system, the consensus protocol of \Architecture is based on \techterm{Proof of Stake (PoS)}. 
    \end{itemize}
    
    \item \textit{Participation in network} 
    \begin{itemize}
        \item The evolution of the \Architecture blockchain is comprised of fixed intervals\footnote{%
                 We envision that the length of an epoch will be measured in the number of blocks. 
              }, called \techterm{epochs}. 
        \item To participate in the network, a node must put up the minimum required stake for that role in a specific epoch
        \item A \Architecture node may participate over multiple epochs. 
    \end{itemize}
    
    \item \textit{Source of randomness}
    \begin{itemize}
        \item \Architecture requires a reliable source of randomness for seeding its pseudo-random number generators. 
        \item The source of randomness enables each seed to be unpredictable by any individual node until the seed itself is generated and published in a fully decentralized manner.
        \item In \Architecture, we use the \techterm{Distributed Random Beacon} (DRB) protocol \cite{hanke2018dfinity} to generate a fully decentralised, reliable source of randomness.
    \end{itemize}
    
    \item \textit{Cryptography Primitives}
    \begin{itemize}
        \item \Architecture requires an aggregatable and non-interactive signature scheme, such as BLS \cite{Boneh:2018:BLS:CompactMF}.
    \end{itemize}
    
    \item \textit{Network Model}
    \begin{itemize}
        \item \Architecture operates on a partially synchronous network with message traverse time bound by $\Delta_{t}$ and the relative processing clock of the nodes bound by $\phi_{t}$.
        \item \Architecture uses a \techterm{Byzantine Fault Tolerant} message routing system, that guarantees message delivery with a high probability. 
    \end{itemize}
    
    \item \textit{Rewarding and Slashing}
    \begin{itemize}
        \item \Architecture requires adequate compensation and slashing mechanics that incentivize nodes to comply with the protocol.
    \end{itemize}
    
    \item \textit{Honest Stake Fraction}
    \begin{itemize}
    \item \Architecture requires \emph{more} than $\frac{2}{3}$ of stake from Collection Nodes, Consensus Nodes, and Verification Nodes to be controlled by honest actors (for each node role separately). We will refer to a group of nodes with \emph{more} than $\frac{2}{3}$ of stake as a \techterm{super-majority}. A super-majorities of honest nodes probabilistically guarantees the safety of the \Architecture protocol. 
    \end{itemize}
\end{itemize}

\subsection{\Architecture's Roles}

\begin{figure}[b!]
\centering
\includegraphics[width=0.7\textwidth]{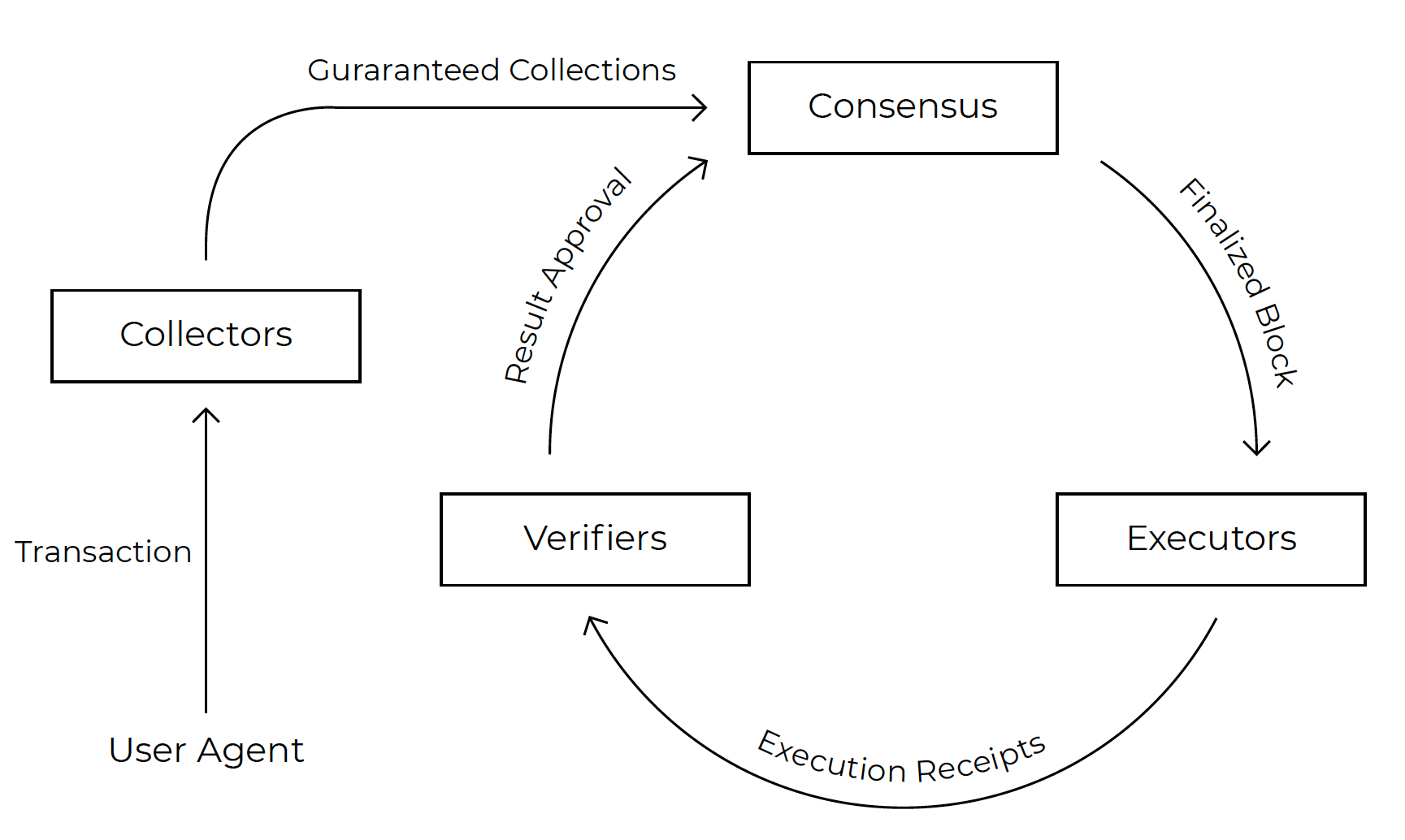}
\caption{An overview of the different roles in \Architecture as well as their interaction. 
        For simplicity, only the message exchange during normal operation is shown.
        Messages for raising or adjudicating slashing challenges are omitted.}
\label{flow:figure_role_overview}
\end{figure}

In \Architecture, roles are the services that the nodes provide to the system. One may assume each role as a specific set of protocols that a node executes. Accordingly, in \Architecture, we have four different roles (i.e., set of protocols), which are:  \textit{Collector Role}, \textit{Consensus Role}, \textit{Execution Role}, and \textit{Verification Role}. We refer to the network nodes that execute the set of respective protocols as \textit{Collector Node}, \textit{Consensus Node}, \textit{Execution Node}, and \textit{Verification Node}. For each role, a minimum stake deposit is required from each of the participating nodes. Hence, single hardware system may host multiple roles in the network by staking for each of them individually. However, \Architecture treats individual roles as if they are independent entities. In other words, each role is staked, unstaked, and slashed independently. The staked nodes associated with the roles are compensated through a combination of block rewards and transaction fees. Any combination of a public key and a role should be unique, and each peer has an independent staked value for each of its roles. We also recommend that multiple roles of a single peer do not share their staking keys for security reasons.

\begin{figure}[b!]
\centering
\includegraphics[width=0.85\textwidth]{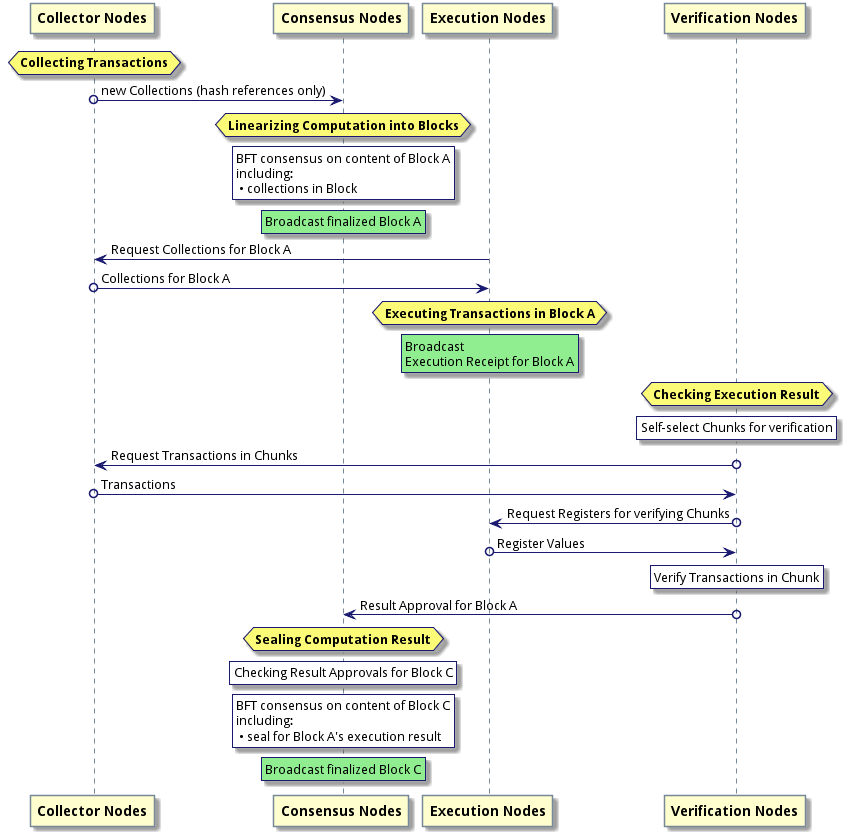}
\caption{%
    The lifecycle of a transaction in \Architecture. The yellow hexagons indicate the start of the individual stages in the \Architecture pipeline, e.g., the block sealing (i.e., Sealing Computation Results). The arrows show the inter-role message exchange between the nodes. Green rectangles correspond to the broadcasting events, in which a message is disseminated to all of the staked nodes. White rectangles represent the execution of the operations of the nodes.
    For the sake of simplicity, we represent the normal flow of the transaction lifecycle and ignored the message exchanges related to adjudicating slashing challenges.
}
\label{flow:figure_transaction_lifecycle}
\end{figure}

We present a detailed description of each node role below.  A high-level illustration of the role interactions is shown in Figure \ref{flow:figure_role_overview}. Figure \ref{flow:figure_transaction_lifecycle} shows an overview of the transaction's lifecycle over considering the different roles of the nodes.

\subsubsection{Collector Role}
For the sake of load-balancing, redundancy, and Byzantine resilience, the Collector Nodes are staked equally and randomly partitioned into clusters of roughly identical size. At the beginning of an epoch, each Collection Node is randomly assigned to exactly one cluster. Each cluster of Collector Nodes acts as a gateway of \Architecture with the external world. In the mature \Architecture, we envision that a cluster will contain somewhere between 20 and 80 Collector Nodes. 

An external client submits their transaction to a Collector Node. Upon receiving a submitted, well-formed \footnote{As we detail later in this paper, a well-formed transaction has all required fields filled properly and contains valid signature(s) from staked account(s) of \Architecture
} transaction, a Collector Node introduces it to the rest of its cluster. The Collector Nodes of one cluster batch the received transactions into so-called \techterm{collections}. Only a hash reference to a collection is submitted to the Consensus Nodes for inclusion in a block. 

Each cluster of the collector nodes generates their collections one at a time. Before a new collection is started, the current one is closed and sent to the Consensus Nodes for inclusion in a block. The collections of a cluster are built collaboratively by each Collector Node sharing the transactions submitted to it with the rest of its cluster, and participating in a (light) consensus protocol with the other Collector Nodes of its cluster. The nodes come to consensus on both the end of the current collection, to start a new one, and on the transactions  included in the new collection. The collection generated as the result of a consensus among the Collector Nodes of a cluster is called a \techterm{guaranteed collection}. 


\subsubsection{Consensus Role}
In \Architecture, the Consensus Nodes maintain the chain of blocks and are responsible for the chain extension by appending new blocks. They receive hash references to the \techterm{guaranteed collection}s that were generated by the Collector Nodes. Furthermore, Consensus Nodes run a \techterm{Byzantine Fault Tolerant} (BFT) consensus algorithm to reach an agreement over the set of \techterm{collections} to be included in the next block. The block of the ordered collection that has undergone the complete BFT consensus algorithm is called \techterm{finalized block}s. In \Architecture, a block specifies\footnote{%
    A block \emph{implicitly} specifies its transactions by referencing collections of transactions.
} the included transactions as well as the other inputs (e.g., the random seed), which are required to execute the computation. It is worth noting that a block in \Architecture does not include the resulting execution state of the block execution. 
  
Consensus Nodes are also responsible for sealing a block. A \techterm{Block Seal} is a commitment to the execution result of a block after it is executed and verified (see \cite{hentschel2019flow} for more details on Block Sealing). Moreover, Consensus Nodes are responsible for maintaining a part of the state of the system related to the stakes of the nodes, receiving and adjudicating the slashing challenges, and slashing faulty nodes. We elaborate on the notion of protocol state on \Architecture in Section \ref{flow:subsec_state}.

\subsubsection{Execution Role}
The Execution Nodes are powerful computational resources that are primarily responsible for scaling the computational power of \Architecture. Execution Nodes execute the finalized blocks generated by the Consensus Nodes. They publish the resulting execution state as an \techterm{Execution Receipt}. The Execution Nodes also need to provide the required information to the Verification Nodes so they can check the execution result. For this purpose, Execution Nodes break the computations of a block into \techterm{chunks}. Each Execution Node publishes additional information about each chunk in its Execution Receipt for the block executed. We detail the block formation process and chunking in Section \ref{flow:sec_consensus} of this paper. 

\subsubsection{Verification Role}
Verification Nodes are responsible for collectively verifying the correctness of the Execution Nodes' published results. With the chunking approach of \Architecture, each Verification Node only checks a small fraction of chunks. A Verification Node requests the information it needs for re-computing the chunks it is checking from the Execution Nodes. A Verification Node approves the result of a chunk by publishing a \techterm{Result Approval} for that chunk, which means that the Verification Node has verified and agrees with the result of the execution of that chunk.

Breaking the verification work into small chunks enables Verification Nodes to check the execution of chunks independently and in parallel. However, as we formally have shown in \cite{hentschel2019flow}, all Verification Nodes together will check all chunks of the executed blocks with an overwhelming probability.


\subsection{States, Staking, and Slashing}
\label{flow:subsec_state}

\textbf{State:} There are two sorts of states in \Architecture blockchain known as the \techterm{execution state} and \techterm{protocol state}. Each type of these states is maintained and updated independently. Both states are represented as key-value stores. While the execution state is maintained and updated by the Execution Nodes, the protocol state is maintained and updated by the Consensus Nodes. 

The execution state contains the register values, which are modified by transaction execution. Although the updates on the execution state of the system are done by the Execution Nodes, the integrity of the updates is governed by \Architecture's verification process \cite{hentschel2019flow}. 

Protocol state, on the other hand, keeps track of the system-related features including, all the staked nodes, their roles, public keys, network addresses, and staking amounts. The protocol state is updated when nodes are slashed, join the system  via staking, or leave the system via un-staking. 
Consensus Nodes publish updates on the protocol state directly as part of the blocks they generate. The integrity of the protocol state is guaranteed by the consensus protocol. We elaborate on updates to the protocol state, affected by the consensus protocol, in Section \ref{flow:sec_consensus} of this paper.

\bigskip
\noindent
\textbf{Staking:} A node in \Architecture is required to deposit some stake in order to run a role. This requires the node to submit a staking transaction. The staking transactions for the next epoch take place before a specific deadline in the current epoch. Once the staking transaction is processed by the Execution Nodes, the stake is withdrawn from the node's account balance and is explicitly recorded in the Execution Receipt. Upon Consensus Nodes sealing the block that contains this staking transaction they update the protocol state affected by this staking transaction, and publish the corresponding staking update in the block that holds the seal.
Staked nodes are compensated through both block rewards and transaction fees and all roles require a minimum stake to formally participate in that role. 

\bigskip
\noindent 
\textbf{Slashing:}
Any staked node of \Architecture can detect and attribute misbehavior to another staked node who committed it. Upon detecting and attributing misbehavior, the node issues a slashing challenge against the faulty node. Slashing challenges are submitted to the Consensus Nodes. The slashing challenge is a request for slashing a staked node due to misbehavior and deviation from the protocol. As the sole entity of the system responsible for updating the protocol state, Consensus Nodes adjudicate slashing challenges and adjust the protocol state (i.e., staking balances) of the faulty nodes accordingly. Based on the result of adjudication, the protocol state (i.e., the stake) of a node may be slashed within an epoch. The slashing update is announced in the subsequent blocks and is effective for all honest nodes as soon as they process the block containing the respective stake update.

\newpage
\section{Preliminaries}
\subsection{Adversarial Model}
\label{flow:subse_adv}
We denote the \textit{honest} nodes (i.e., non-Byzantine nodes) as the ones that follow the description of the \Architecture protocols. We call a node \textit{Byzantine} if it deviates from any of \Architecture's protocols at any arbitrary point. We also presume the non-responsiveness of a staked node, to messages and queries, to be a Byzantine act. 
 
\begin{definition}{\textbf{Effective Votes}}\\
During the consensus among a group of nodes, we consider the \textbf{effective votes} of the nodes as the overall staked fraction of the nodes who vote positively in favor of the consensus proposal. The fraction is taken over the entire stakes of the nodes in that group. 
\end{definition}

For the Collector Role, Consensus Role, and Verification Role, we assume that \emph{more than $\consThr$} of the accumulated stake of each role belong to the honest nodes, and the rest is owned by the Byzantine nodes. For example, more than $\consThr$ of the stakes of the Collector Nodes belong to the honest ones and less than $\frac{1}{3}$ of their overall stakes are owned by the Byzantine ones. 
\subsection{\consensus}
\label{flow:subsec_consensus}
\consensus \cite{HotStuff:2019:ACM, HotStuff:2018} is a distributed Byzantine Fault Tolerant (BFT) consensus algorithm. In this subsection, we present a summary of the basic \consensus protocol. However, we scope out the optimization improvements which are presented in the \consensus proposal for sake of brevity, and refer the interested readers to \cite{HotStuff:2018}. 

\consensus assumes the nodes as state machines that hold the same shared replicated data (e.g., the blockchain ledger), and aim to reach consensus over the transition of the next state of the replicated data (e.g., the set of the transactions to be included into the next block). 
\consensus is live  under partially-synchronous network conditions. It is safe if less than one third of the consensus nodes' total stake is controlled by Byzantine actors.  

The protocol progresses in rounds, i.e., each node increments a local counter when reaching consensus, or a timeout. For sake of liveness, the nodes double their timeout interval each time they move to a new round as the result of a time out. 
In each round, a unique consensus node assumes the role of the leader. Nodes can consistently identify the leader of each round by invoking a deterministic function locally on the round number. The leader advances the consensus via a three-phase commit protocol of \techterm{prepare}, \techterm{pre-commit}, and \techterm{commit}. The consensus at each round starts with a block proposal by the leader, and ends with reaching a consensus over the proposal or a timeout. In \Architecture, we assume that moving from one phase to another phase of \consensus requires a minimum effective vote of $\frac{2}{3}$ in favor of progressing with the leader's proposal. The leader collects and aggregates the votes and broadcasts the aggregated signature to the entire system as a proof of moving to the next phase. 
Hence, each node can track and verify the correctness of the consensus progress in each phase by confirming that there are at least $\frac{2}{3}$ effective votes. 

The correctness of the \consensus protocol is formally proven via the safety and liveness theorems \cite{HotStuff:2018}. The safety implies that all honest nodes will eventually commit the same state transitions (i.e., blocks) in the same order.  The liveness property of \consensus guarantees that, after a Global Synchronization Time has passed,  there is a bounded interval that enables an honest leader to advance the round towards reaching a consensus. Furthermore, \consensus provides \techterm{deterministic finality}, i.e., the protocol guarantees that there will be no forks in the chain of committed state transitions. This essentially means that the canonical chain of replicated state transitions (e.g., the ledger of blocks) solely acts as an append-only chain. Hence, in contrast to many existing consensus protocols (including Bitcoin's Proof-of-Work (PoW)), chain reorganizations do not occur in \consensus\footnote{Provided that the stake controlled by Byzantine actors is strictly less than $\frac{1}{3}$.}. 

\bigskip
\noindent
The advantages of using \consensus compared to the existing BFT leader-based counterparts include
\begin{itemize}
    \item Having $\nodesNum$ nodes participating in the consensus, the communication complexity of \consensus is $O(\nodesNum)$, which makes switching of faulty or non-responsive leaders relatively easy. 
    \item The proposal of a failed or timeout leader has the potential to be resumed by the leader of the next round. 
    \item Preferring safety over the liveness, i.e., the safety over \consensus is guaranteed regardless of the underlying synchrony model. However, for the protocol to guarantee its liveness and the progress, the Global Synchronization Time should be passed. 
    \item Ability to operate under partially-synchronous network conditions. 
    \item It has the so-called \textit{responsiveness} property, i.e., an honest leader advances the protocol to the consensus in a time that is bounded by the message transmission delay, i.e., $\Delta_{t}$.
    \item It has deterministic finality. 
\end{itemize}
\subsection{Distributed Random Beacon}
\label{flow:subse_drb}
In \Architecture, we utilize the Distributed Random Beacon (DRB) protocol \cite{hanke2018dfinity} among a subset of nodes to generate an unbiased and verifiable source of randomness in a fully decentralized manner. By unbiased, we mean that no adversarial party can manipulate the source of randomness towards its desired distribution. By verifiability, we mean that once the source of randomness is determined, it is verifiable against the correct execution of the protocol. As a decentralized protocol, DRB is executed collaboratively among a subset of nodes as an interactive multi-party protocol.

In \Architecture, we consider DRB as a 2-phase protocol consisting of three polynomial time protocols: \texttt{Setup}, \texttt{RandomGenerator}, and \texttt{Verify}. The first phase of DRB is the setup phase that happens only once among the participants. The second phase of DRB is the random generation phase, which is a repetitive operation, i.e., it may happen many times within the lifetime of the system. Each time the participants enter the random generation phase, they generate a random string that is denoted as the \textit{\entropy}. In this paper, we call every single execution of the random generation phase of DRB a \textit{DRB round}. 
\begin{itemize}
    \item $(V_{G}, \texttt{sk}_{i})  \leftarrow \texttt{Setup}(P_{i}, 1^{\lambda})$: \texttt{Setup} is a probabilistic decentralized \textit{protocol} where each participating party $i$ receives its private key, i.e., $\texttt{sk}_{i}$, as well as a public verification vector, i.e., $V_{G}$. We elaborate on these in the rest of this section. 
    \item $r \leftarrow \texttt{RandomGenerator}(\texttt{sk}_{i}, \texttt{proto(b)})$: \texttt{RandomGenerator} is a deterministic decentralized \textit{protocol} where parties collaboratively generate a fresh \entropy. For each participating party in this protocol, the inputs are the party's private key (i.e., $\texttt{sk}_{i}$) and a  \textit{proto block} (i.e., $proto(b)$). The output of the protocol is the fresh \entropy, i.e., $r$. We elaborate on the proto block notion in Section \ref{flow:sec_consensus} of this paper. 
    \item $\texttt{Verify}(r, proto(b), V_{G}) = \texttt{True}/\texttt{False}$: \texttt{Verify} is a deterministic protocol that is executed locally by any external or internal party to DRB. Knowing a generated \entropy (i.e., $r$), its associated proto block (i.e., $proto(b)$) and the public verification vector of the DRB (i.e., $V_{G}$), one can deterministically verify the correctness of the protocol execution. 
\end{itemize}

\bigskip\noindent
In addition to being unbiased, DRB provides the following properties: 
\begin{itemize}
    \item \textbf{Unpredictability:} Denoting the security parameter of the system by $\lambda$, before the execution of a new round of DRB, for each probabilistic polynomial-time predictor $A$, there exists a negligible function $negl(\lambda)$, such that:
    \begin{align*}
        Pr[A(r_{1}, r_{2}, r_{3}, ..., r_{x-1}, proto(b_{1}), proto(b_{2}), proto(b_{3}), ..., proto(b_{x-1}), proto(b_{x})) = r_{x}] \\ 
        \leq negl(\lambda)
    \end{align*}
    where $r_{i}$ is the output of the $i^{th}$ execution of DRB, and $proto(b_{i})$ is the associated proto block to the $r_i$. In other words, given all the generated entropies as well as their associated proto blocks to the predictor $A$, it is not able to predict the output of the next round before its execution unless with a negligible probability in the security parameter of the system. 
    
    \item \textbf{Verifiability:} Given the generated \entropy of a round, its corresponding proto block, as well as some public metadata of the protocol, any external party to the protocol can verify the generated \entropy. 
\end{itemize}

\subsubsection{DRB Random Generation Phase (Threshold Signatures)}
The main idea of DRB is based on the existence of a deterministic and unique non-interactive \textit{threshold signature} protocol. BLS signature primitive \cite{Boneh:2018:BLS:CompactMF} provides all these required properties.  
A threshold signature is a tuple of four polynomial-time algorithms known as \texttt{(Gen, Sign, Recover, Verify)}, which is defined over the set $G$ of $n$ parties (i.e., $G = \{P_{1}, P_{2}, ..., P_{n}\}$) and is identified by two parameters: $t$ and $n$. An $(t,n)$-threshold signature enables the set of $n$ parties to collaboratively generate an \textit{aggregated signature} over a message $m$ using a distributed secret key in a decentralized manner. The threshold signature enables any entity to efficiently verify a valid signature over $m$ against the distributed secret key. The verification is solely done by verifying the aggregated signature, without requiring the individual party's signatures.   
\begin{itemize}
    \item $(V_{G}, \texttt{sk}_{1}, \texttt{sk}_{2}, ..., \texttt{sk}_{n})  \leftarrow \texttt{Gen}(1^{\lambda})$: \texttt{Gen} is a probabilistic polynomial-time algorithm that on receiving the security parameter, it implicitly generates a secret key $\texttt{sk}_G$ and distributes individual key shares over the participating parties i.e., $\texttt{sk}_{i}$ for the $i^{th}$ party. $\texttt{sk}_G$ remains secret and not explicitly computed although it is jointly generated by all parties. Moreover, a public verification vector $V_{G}$ is generated. $V_{G}$ enables parties to recover each others' public key shares. By the public key of the $i^{th}$ party, we mean the public key $\texttt{pk}_i$ associated with $\texttt{sk}_{i}$. $V_{G}$ also contains the public key of the group of parties, i.e., $\texttt{pk}_{G}$ associated with $\texttt{sk}_G$. 
    \item $\sigma_{i}  \leftarrow \texttt{Sign}(\texttt{sk}_{i}, m)$: \texttt{Sign} is a deterministic polynomial-time algorithm that is executed by each party individually. On receiving a message $m$ and the private key of the $i^{th}$ party (i.e., $\texttt{sk}_{i}$), $\texttt{Sign}(\texttt{sk}_{i}, m)$ generates a signature $\sigma_{i}$ on the message $m$ that is attributable to the $i^{th}$ party. $\sigma_{i}$ is also commonly called the \textbf{threshold signature share} of the $i^{th}$ party. \texttt{Sign} is the same signature function defined in the BLS signature scheme.
    \item $\sigma  \leftarrow \texttt{Recover}(P_{i_1}, P_{i_2}, ..., P_{i_{t+1}}, \sigma_{i_1}, ..., \sigma_{i_{t+1}})$: \texttt{Recover} is a deterministic polynomial-time algorithm that is executed by any entity either internal or external to the group of parties. On receiving the distinct threshold signature shares $\sigma_{i_k}$ and the list of corresponding distinct parties, \texttt{Recover} combines all the signature shares and recovers the group signature, i.e., $\sigma$. Having any combination of more than $t$ signature shares of the parties over the same message $m$, \texttt{Recover} generates a group signature $\sigma$ that is attributable to the group secret key, i.e., $\texttt{sk}_{G}$.
    \item $\texttt{Verify}(\sigma, \texttt{pk}_{G}, m) = \texttt{True}/\texttt{False}$: \texttt{Verify} is a deterministic polynomial-time algorithm that is executed by any entity either internal or external to the group of parties. On receiving a recovered threshold signature (i.e., $\sigma$) on a message $m$, it returns \texttt{True} if the signature $\sigma$ was generated using the group secret key (i.e., $\texttt{sk}_{G}$), and returns \texttt{False} otherwise. \texttt{Verify} is the same signature function defined in the BLS signature scheme.
\end{itemize}

\noindent
While \texttt{Sign}, \texttt{Recover} and \texttt{Verify} are non-interactive and relatively inexpensive operations, \texttt{Gen} requires to run an interactive and relatively slow protocol by all parties. Keeping the performance factor in mind, \texttt{Gen} is separated from the repetitive DRB random generation phase in \Architecture. \texttt{Gen} serves as the \texttt{Setup} phase while the tuple  (\texttt{Sign}, \texttt{Recover}, \texttt{Verify}) serves as the random generation phase. This is possible because the random generation allows generating repetitive signatures using $\texttt{sk}_G$ without explicitly reconstructing it, and therefore protecting its confidentiality.
The DRB random generation phase is started every time the nodes need to generate a new \entropy. Protocol \ref{protocol:DRBRandomGenerator} shows a DRB random generation round for a specific proto block, i.e., $proto(b)$. To generate a new \entropy, each node $i$ computes and broadcasts a threshold signature share over $proto(b)$. Upon receiving at least $t+1$ distinct threshold signature shares $\{\sigma_{b,1}, \sigma_{b,2}, \sigma_{b,3}, ..., \sigma_{b,t+1}\}$ over the proto block, any party can aggregate the signatures into a threshold signature $\sigma_{b}$, using the function \texttt{Recover}. If the signature $\sigma_{b}$ is correctly formed, it is verifiable against the public key of the group $G$, i.e., $\texttt{Verify}(\sigma_{r}, proto(b), \texttt{pk}_{G}) = \texttt{True}$. The \entropy of this DRB random generation round is $\sigma_{b}$. The deterministic 
digest of $\sigma_{b}$ is used to seed the pseudo-random generators during the current round.

\medskip
\begin{protocol}{DRB Random Generator\label{protocol:DRBRandomGenerator}}
\textit{Inputs:} For each $i \in [1,n]$, party~$P_i$ holds the inputs of $proto(b)$ and $\texttt{sk}_{i}$. $H$ is a cryptographic hash function.
\sbline
\textit{Goal:} Parties jointly compute a fresh \entropy, $r$, based on $proto(b)$.
\sbline
\textit{The protocol:}
\begin{enumerate}
    \item
    Each party $P_{i}$ generates $\sigma_{b,i} \leftarrow \texttt{Sign}(proto(b),\texttt{sk}_{i})$, and broadcasts $\sigma_{b,i}$ to all other parties.

    \item
    Any internal or external party then captures the broadcasted $\sigma_{b,j}$

    \item
    Upon having at least $t+1$-many $\sigma_{b,.}$ from $t+1$ distinct parties, each party recovers the aggregated threshold signature $r=\sigma_{b} \leftarrow \texttt{Recover}(P_{i_1}, ..., P_{i_{t+1}}, \sigma_{b,i_1}, ..., \sigma_{b,i_{t+1}})$. 

    \item
    Having $r=\sigma_{b}$, any party extracts a fresh seed as $H(r)$
\end{enumerate}
\end{protocol}

In \Architecture, we use \cite{boldyreva2003threshold} as the underlying threshold signature scheme for the DRB, which is non-interactive and provides unique signatures. By non-interactive, we mean that the protocol is modeled by a single round of communication. Uniqueness is an inherited property from the BLS scheme, and it means that there exists a unique aggregated threshold signature for each pair of $(\texttt{pk}_{G}, m)$. Hence, the aggregated threshold signature is deterministic and is always the same regardless of the subset of collected threshold signature shares over the same message. The non-interactiveness feature is for sake of improvement on the communication complexity of the DRB protocol over several instances (i.e., rounds) of execution. The uniqueness is vital to guarantee a consistent \entropy independent of the subset of threshold signature shares, i.e., every valid subset of $t+1$-many threshold signature shares results in the same aggregated threshold signature.

\subsubsection{DRB Setup Phase (Distributed Key Generation)}
Distributed Random Beacon (DRB) setup phase is to generate the keys for the threshold signature protocol. As shown by Protocol \ref{protocol:DRBSetup}, the setup phase of DRB in \Architecture is mainly the \texttt{Gen} function of the threshold signature. During this phase, the parties of the group $G$ collaboratively execute a \textit{Distributed Key Generation} protocol (DKG) \cite{gennaro1999secure} as the decentralized implementation of \texttt{Gen} function. The DKG implicitly generates a group secret key $\texttt{sk}_G$. As the result, each party $P_{i}$ receives the public verification vector $V_{G}$ as well as its individual private key $\texttt{sk}_{i}$. The secret key $\texttt{sk}_G$ is generated using entropy provided by all participating parties in an unbiased manner. It remains unknown to all participants although it can be reconstructed by any party receiving more than $t$ shares of the secret key.

The DKG implemented in \Architecture is a variant of \textit{Pedersen}'s protocol called \textit{Joint-Feldman} \cite{gennaro1999secure} protocol. It is a Discrete Logarithm-based protocol and mainly constitutes of $n$ parallel executions of the \textit{Feldman verifiable secret sharing} (VSS) protocol with each party acting as a leader in exactly a single instance. Although it has been proven that Joint-Feldman protocol does not guarantee a uniform distribution of the secret key $sk_G$, the security of the DKG in \Architecture is based on the hardness of the Discrete Logarithm Problem, which is not weakened by the distribution bias.

\medskip
\begin{protocol}{DRB Setup\label{protocol:DRBSetup}}
\textit{Inputs:} For each $i \in [1,n]$, party~$P_i$ holds its index in the set as well as the system's security parameter, $\lambda$.
\sbline
\textit{Goal:} Parties jointly compute a public verification vector $V_{G}$ as well as their  threshold signature private keys, i.e., $\texttt{sk}_{i}$ for $P_{i}$.
\sbline
\textit{The protocol:}
\begin{enumerate}
    \item
    Each party $P_{i}$ invokes an instance of DKG protocol as specified in \cite{gennaro1999secure}, and obtains:
    
    $(V_{G}, \texttt{sk}_{i}) \leftarrow \texttt{DKG}(P_{i}$, $1^{\lambda}$)
\end{enumerate}
\end{protocol}

\subsection{Distributed Random Beacon Setup in \Architecture}
We assume the setup phase of DRB happens once every epoch to prepare the DRB protocol keys. During the setup phase, a protocol-selected subset of $\consensusNum$ Consensus Nodes jointly execute the DRB setup protocol. Through executing the setup protocol, each node $i$ generates its \textit{DRB private key} 
$\texttt{sk}_i$ 
for the threshold signature as well as a public verification vector $V_{G}$.

As explained in Section \ref{flow:subse_drb}, $(n,t)$ are the parameters of the threshold signature scheme. An adversary who corrupts up to $t$ parties is not able to forge a valid threshold signature. It is also required that at least $t+1$ parties act honestly to guarantee the liveness of the protocol. In order to satisfy both properties with the high probability, i.e unforgeability and liveness, we compute $t$ as follows.
\begin{equation}
    t = \floor{\frac{\consensusNum-1}{2}}
    \label{flow:eq_t}
\end{equation}
In \Architecture, the size of the DRB committee $\consensusNum$ is tuned to ensure that unforgeability and liveness are guaranteed with a high probability in the security parameter of the system (i.e., $\lambda$). However, $\consensusNum$ also is chosen small enough to preserve the efficiency of the DRB execution with respect to the operational complexities.

\newpage
\section{Collection Formation}
\label{flow:sec_collection}
\subsection{Overview}
In \Architecture, \techterm{collection formation} denotes the process which starts with a user agent submitting a transaction to the Collector Nodes and ends when a guaranteed collection is submitted to the Consensus Nodes. The Collector Nodes are partitioned into clusters. Each transaction is assigned to a certain cluster based on its transaction hash. A user agent submits a transaction to some Collector Nodes in the responsible cluster. On receiving a transaction, the Collector Nodes of the cluster broadcast the transaction among each other. Collector Nodes of a cluster continuously form consensus over the set of transactions contained in the collection under construction. As a result of the consensus, a collection grows over time. Once one of two conditions is met: either the collection size reaches a threshold or a pre-defined time span has passed, the collection is closed and submitted to the Consensus Nodes for inclusion in a block. 


\subsection{Cluster Formation}

\noindent
In \Architecture architecture, the stake is a measure of the nodes' accountability. In specific for the Collector Nodes, the workload accountability of processing the submitted transactions and hence their compensation is a monotonically increasing function of their stake \textit{deposit per processed transaction}. To make all the Collector Nodes equally accountable, we desire a setup where the stake deposit per processed transaction is similar for all the Collector Nodes. This setup with similarly staked Collector Nodes results in a similar workload distribution among them on processing the submitted transactions. This setup stands against the non-uniform stake distribution of the Collector Nodes, which results in a non-uniform workload of processing the submitted transactions among them. 

Despite our equally staked Collector Nodes principle, the user agents submitting their transactions to an arbitrary Collector Node of their choice results in a biased distribution of the workload and its associated rewards. It is trivial that over the time the user agents may establish a relationship with Collector Nodes, e.g., running the client software provided by the Collector Nodes, submitting to the Collector Node that provides a better quality of service, etc. This, however, leads to a heterogeneous load distribution on the Collector Nodes and degrades the decentralization of the system. For example, a Collector Node may provide significantly higher utilization than the others, attract more rewards and essentially starve other Collector Nodes of income. 

To design a system where the stake deposit per processed transaction is comparable for all collectors, \Architecture introduces the notion of \textit{clusters}. Several Collector Nodes are grouped into the same cluster to collect and process the same set of transactions. In other words, instead of having a pool of transactions where each Collector Node arbitrarily selects transactions to process, there is a one-way deterministic assignment between each transaction and a cluster of the Collector Nodes using the \entropy. Clustering of Collector Nodes is done at the beginning of each epoch, e.g., once every week. The number of clusters in \Architecture is a protocol parameter denoted by $\clusterNum$. As the result of clustering, the Collector Nodes are randomly partitioned into $\clusterNum$ clusters. Clustering is done in a way that the size of each two different clusters varies by at most a single node. The assignment is nevertheless verifiable, i.e., each node is able to run the clustering algorithm offline and reach the same cluster assignment for the Collector Nodes as every other node. As we detail in the rest of this section, this enables the user agents to efficiently map their signed transactions to the cluster of the Collector Nodes that is responsible for processing them.

The clustering is done by each Collector Node running the cluster assignment algorithm as illustrated by Algorithm \ref{flow:alg_clustering}. The inputs to this algorithm are the list of all Collector Nodes' public keys (i.e., $\collector$), the number of clusters (i.e., $\clusterNum$), and the \entropy (i.e., $\randomBeacon$), which is generated by the Distributed Random Beacon protocol \cite{hanke2018dfinity} (see Section \ref{flow:subse_drb}). We denote the cardinality of $\collector$ as $\collectorsNum$, which corresponds to the number of Collector Nodes in the system. The output of the algorithm is $\cluster$, which is a map from the public keys of the Collector Nodes to their assigned clusters, i.e., for every $\texttt{pk} \in \collector$, element $\cluster[\texttt{pk}]$ is the Collector Node's assigned cluster id. 

The clustering is done by deriving a seed $s$ from the \entropy $r$ solely for the clustering of Collector Nodes (Algorithm \ref{flow:alg_clustering}, Line \ref{flow:alg_clustering_seed}). The seed is derived in a deterministic manner and has the same entropy as $r$.
The derived seed is then being utilized by the Fisher-Yates shuffling algorithm \cite{fisher1943statistical}. 
On receiving the seed $s$ and the list of Collector Nodes' public keys $\collector$, the Fisher-Yates algorithm turns $\collector$ into a pseudo-random permutation of Collector Nodes' public keys, denoted by $\pi_{c}$. 
Note that as $\pi_{c}$ is a permutation of $\collector$, it is of the same cardinality as $\collector$, i.e., $|\collector| = |\pi_{c}| = \collectorsNum$ (Algorithm \ref{flow:alg_clustering}, Line \ref{flow:alg_clustering_shuffeling}). 

Once the permutation $\pi_{c}$ is determined, the clustering algorithm partitions the Collector Nodes into $\clusterNum$ clusters. It does that by clustering the first $\clusterSize$ Collector Nodes into the cluster number $0$, the second $\clusterSize$ Collector Nodes into the cluster number $1$, and similarly the $i^{th}$ sequence of the Collector Nodes of size $\clusterSize$ into the cluster number $i-1$. $\clusterSize$ is the size of each cluster and is determined as $\clusterSize := \floor{\frac{\collectorsNum}{\clusterNum}}$ (Algorithm \ref{flow:alg_clustering}, Lines \ref{flow:alg_clustering_clustering_begin}-\ref{flow:alg_clustering_clustering_end}).

In case the number of Collector Nodes is a multiple of the number of clusters, i.e., $\floor{\frac{\collectorsNum}{\clusterNum}} = \frac{\collectorsNum}{\clusterNum} = \ceil{\frac{\collectorsNum}{\clusterNum}} = \clusterSize$, clustering algorithm stratifies the Collector Nodes based on their position in the permutation $\pi_{c}$ into $c$ clusters of size $\clusterSize$. Otherwise, if the number of Collector Nodes is \textit{not} a multiple of the number of clusters, clustering based on Lines \ref{flow:alg_clustering_clustering_begin}-\ref{flow:alg_clustering_clustering_end}, results in some Collector Nodes to be leftover (Specifically, the number of leftover Collector Nodes will be less than $\clusterNum$). In that situation, the clustering algorithm does not create an extra cluster to accommodate the leftovers. It rather distributes the leftovers among the existing clusters by adding the $i^{th}$ Collector Node of the leftovers to the $i^{th}$ cluster. Size of the leftovers in this case is $\collectorsNum \, \texttt{mod} \, \clusterNum$, i.e., the remainder of $\collectorsNum$ divided by $\clusterNum$. Hence, clustering in this manner results $\collectorsNum \, \texttt{mod} \, \clusterNum$ clusters of $\clusterSize + 1$ Collector Nodes, and $c - \collectorsNum \, \texttt{mod} \, \clusterNum$ clusters of $\clusterSize$ Collector Nodes (Algorithm \ref{flow:alg_clustering}, Lines \ref{flow:alg_clustering_redistribution_begin}-\ref{flow:alg_clustering_redistribution_end}). Having $\collectorsNum$ Collector Nodes in the system, Algorithm \ref{flow:alg_clustering} has both an asymptotic time and memory complexity of $O(\collectorsNum)$, and does not impose a communication overhead to the system. \\

\begin{algorithm}[H]
{
\KwIn{\\
    \quad $\collector$: List; element $\collector[i]$ is public key of $i^{th}$ Collector Node\\
    \quad $\clusterNum$: System-wide constant unsigned integer; number of the clusters\\
    \quad $\randomBeacon$: byte array; \entropy
    }
\KwOut{\\
        \quad $\cluster$: Map; element $\cluster[\texttt{pk}]$ is the cluster id of Collector Node with public key of $\texttt{pk}$
      }
    
\BlankLine
    \tcp{Deriving a random seed from $r$ for clustering Collector Nodes}
    $s$ := $H(\texttt{`collector'}||\texttt{`cluster'}||\randomBeacon)$\;
    \label{flow:alg_clustering_seed}
    \tcp{Shuffling the collectors list}
    $\pi_{c} :=$ \texttt{FisherYates}$(s, \collector)$\; 
    \label{flow:alg_clustering_shuffeling}
       
    \tcp{$\clusterSize$ defines the size of clusters}
    $\clusterSize := \floor{\frac{\collectorsNum}{\clusterNum}}$\;
    \label{flow:alg_clustering_clustering_begin}
    \tcp{$i$ keeps current cluster's id}
    $i := 0$\;
    \tcp{$j$ keeps current Collector Node's index in $\pi_{c}$}
    $j := 0$\;
    \tcp{stratifying Collector Nodes into $\clusterNum$ clusters of size $\clusterSize$}
    \While{$j < \clusterNum \times \clusterSize$}
    {
        \tcp{adding $j^{th}$ Collector Node to cluster $i$}
        $\cluster[\pi_{c}[j]] := i$\;
        \tcp{moving to the next Collector Node}
        $j++$\;
        \uIf{$j \, \texttt{mod} \, \clusterSize = 0$}
        {
            \tcp{current cluster has reached size $\clusterSize$}
            \tcp{moving to the next cluster}
            $i++$\;
        }
    }
    \label{flow:alg_clustering_clustering_end}
    $i := 0$\;
    \label{flow:alg_clustering_redistribution_begin}
    \While{$j < \collectorsNum$}
    {
        \tcp{adding $j^{th}$ Collector Node to cluster $i$}
        $\cluster[\pi_{c}[j]] := i$\;
        \tcp{moving to the next cluster}
        $i++$\;
        \tcp{moving to the next Collector Node}
        $j++$\;
    }
    \label{flow:alg_clustering_redistribution_end}
\Indp
\Indm
\caption{ClusterAssignment}
\label{flow:alg_clustering}
}
\end{algorithm}

\subsection{Transaction submission}

Listing \ref{message:tx} represents the structure of the \Architecture's transactions that the user agents submit to the Collector Nodes. In this listing, \texttt{Script} corresponds to the content of the transaction. The transaction's content manipulates the execution state of \Architecture, and requires a transaction fee to be processed. \texttt{PayerSignature} corresponds to the signature created by the account that is paying the transaction fee (e.g., the gas fee). The \texttt{ScriptSignature} of the transaction corresponds to the signatures of the execution state owners, which grant the transaction permission to manipulate their execution state. A transaction may manipulate several parts of the execution state and hence may require the signatures of several execution state owners. \texttt{ReferenceBlockHash} points to the hash of a previous block, which is used to provide a deadline for the transaction to be processed. Each submitted transaction must be processed within a limited window of blocks, e.g., the height difference between the block that the transaction's \texttt{ReferenceBlockHash} points to and the block that the transaction appears in should be less than a predefined parameter. For example, assume that the window size for the transactions is $10$ blocks and a transaction is pointing to a block at height $1000$ in its \texttt{ReferenceBlockHash}. This means that the transaction can only be processed between blocks of height $1001$ and $1010$. Otherwise, the transaction is invalid.

Let $h$ denote hash of the \texttt{SignedTransaction}, interpreted as an \emph{unsigned integer}. To submit its transaction, the user agent determines the cluster of the Collector Nodes that are responsible for collecting its transaction. The collection determination is done as shown by Equation \eqref{flow:eq_collection_assignment}, where $\clusterNum$ corresponds to the number of clusters in the system. The index of the cluster, to which the transaction with hash $h$ should be submitted to is
\begin{equation} \label{flow:eq_collection_assignment}
    c_{h} = h \hspace{3pt} \texttt{mod} \hspace{3pt} \clusterNum \, .
\end{equation}
Here, $\texttt{mod}$ is the modulus operation that determines the remainder of the integer-division of the left-side operand over the right-side operand.

In \Architecture, the assignment of Collector Nodes to clusters is easily computed in a deterministic fashion by executing Algorithm \ref{flow:alg_clustering}. Hence, a user agent can efficiently map its signed transaction to a cluster of the Collector Nodes using Algorithm \ref{flow:alg_clustering} and Equation \eqref{flow:eq_collection_assignment}. The transaction submission is done by the user agent sending its signed transaction to the cluster with index $c_{h}$ that is driven by Equation \eqref{flow:eq_collection_assignment}. By sending a transaction to a cluster, we mean sending to at least one of the Collector Nodes belonging to that cluster. However, for sake of fault tolerance in the presence of Byzantine Collector Nodes as well as (network) failures, a user agent may send the same transaction to several Collector Nodes of the assigned cluster. In \Architecture, we leave the number of Collector Nodes that a user agent needs to submit its transaction as a user-dependent decision. A user with a strong set of utilities may prefer to send all or only its important transactions to the entire cluster, while another user may prefer to submit it only to a single node of the assigned cluster. Upon receiving a transaction from a user agent, the Collector Node checks whether the transaction is submitted to the correct cluster according to Equation \eqref{flow:eq_collection_assignment} as well as whether the transaction is well-formed. A transaction is well-formed if it contains all the fields as shown by Listing \ref{message:tx} as well as valid \texttt{ScriptSignature}(s) and \texttt{PayerSignature} by the registered accounts of \Architecture. If the transaction is well-formed as well as submitted to the right cluster, the Collector Node broadcasts the signed transaction to the entire cluster. 


\noindent
\begin{minipage}{\linewidth}
\begin{lstlisting}[language=protobuf2,style=protobuf, caption={User agent broadcasts the transaction to multiple Collector Nodes of the assigned cluster.}, label={message:tx}]
message SignedTransaction {
 bytes Script;
 bytes PayerSignature;
 repeated Signature ScriptSignature;
 bytes ReferenceBlockHash;
}
\end{lstlisting}
\end{minipage}

\subsection{Collection Formation\label{flow:subsec_collection}}
In \Architecture, the term \textit{collection} refers to an ordered list of one or more \textbf{hashes} of signed transactions. 
Collector Nodes form collections by coming to consensus on the ordered list of transaction hashes. As discussed in Section \ref{flow:subsec_consensus}, \Architecture uses \consensus \cite{HotStuff:2018, HotStuff:2019:ACM} as the consensus protocol among the Collection Node clusters, for them to form collections. However, any BFT consensus protocol with deterministic finality would apply to the architecture. In each round of consensus, the selected leader either resumes an uncompleted or time-outed proposal of the previous round or makes a new proposal. A proposal is either on appending a list of the signed transactions' hashes to the current collection or on closing the current collection and starting a new empty collection. For each phase of the consensus to proceed a minimum effective vote of $\consThr$ on the leader's message is required. For a Collector Node to vote in favor of an append proposal, the following conditions must all be satisfied:
\begin{itemize}
    \item The Collector Node has received all the signed transactions that their hashes represent are reflected in the append proposal.
    \item Each of the transaction hashes represented in the append proposal should be well-formed and signed by at least one of the Collector Nodes of the same cluster. 
    \item Appending the transactions from the proposal to the collection should not result in duplicate transactions. 
    \item There should not be common transactions between the current collection under construction and any other collection that the cluster of the Collector Node has already guaranteed.
\end{itemize}
Once a message has the minimum effective vote of $\consThr$, the leader aggregates the signatures associated with the individual votes which advances the consensus process to the next phase. In this way, every individual Collector Node can verify the correctness of the consensus process in a fully decentralized manner.

Initially, a new empty collection is created as a shared replicated state in every Collector Node. Upon reaching a consensus over an append operation, the signed transactions' hashes are appended to the collection. Once the collection size reaches a protocol's predefined threshold, a non-Byzantine leader of the subsequent rounds propose closing the collection. Consensus over the closing a collection is done by following the HotStuff protocol. The structure of a Guaranteed Collection is presented in Listing \ref{message:collection}, where \texttt{CollectionHash} is the hash of the closed collection. \texttt{ClusterIndex} is the id of the cluster (see Algorithm \ref{flow:alg_clustering}) that has generated the collection, reached a consensus over and closed it. \texttt{AggregatedCollectorSigs} holds the aggregated signatures of the Collector Nodes that have guaranteed the collection. For the \texttt{GuaranteedCollection} to be considered as valid, \texttt{AggregatedCollectorSigs} should contain at least $\consThr$ effective votes of the Collector cluster. 
We call each of the Collection Nodes that participate in making a guaranteed collection, a \textbf{guarantor} of that collection. By singing a collection, a guarantor attests to the followings:
\begin{itemize}
    \item All the transactions in the collection are well-formed. 
    \item They will store the entire collection including the full script of all transactions.
\end{itemize}
In \Architecture, we consider a guaranteed collection to be an immutable data structure. The guaranteed collection is broadcasted by the guarantors to all Consensus Nodes for inclusion in a block. 


\noindent
\begin{minipage}{\linewidth}
\begin{lstlisting}[language=protobuf2,style=protobuf, caption={Each Collector Node broadcasts the guaranteed collection reference to all the Consensus Nodes.}, label={message:collection}]
message GuaranteedCollection {
    bytes   CollectionHash;
    uint32  ClusterIndex;
    Signature AggregatedCollectorSigs;
}
\end{lstlisting}
\end{minipage}

\subsection{Liveness}
We introduce Lemma \ref{lemma:safety_1} before presenting the safety theorem of the collection formation process, and directly utilize it as part of the proof for the safety of the collection formation process.  
\begin{lemma}
\label{lemma:safety_1}
Given a guaranteed collection that is generated by a set of guarantors in a cluster with the following conditions:
\begin{itemize}
    \item \textbf{Network Model:} partially synchronous network with message traverse time-bound by $\Delta_{t}$ and the relative processing clock of the nodes bound by $\phi_{t}$
    \item  \textbf{Message Delivery:} a fault-tolerant message routing functionality that guarantees the message delivery with high probability 
    \item \textbf{Stake Distribution:} all the Collector Nodes are equally staked
    \item \textbf{Honest Behavior:} the honest actors follow the protocols as specified, and maintain their availability and responsiveness in the system over time
    \item \textbf{Honest Lower-bound:} the guaranteed collection has at least one honest guarantor
\end{itemize}
The guaranteed collection is always available and recoverable in the system.   
\end{lemma}

\noindent
\textbf{Proof of Lemma \ref{lemma:safety_1}:}
As detailed in Section \ref{flow:subsec_collection}, the guarantors of a guaranteed collection maintain the \textit{original copy} of it, and only broadcast the \textit{collection reference} that is presented as a \textit{GuaranteedCollection} to the Consensus Nodes. In other words, a guaranteed collection is supposed to be replicated on all of its guarantors, i.e., the Collector Nodes that signed it. By contradiction, assume that the original content of a guaranteed collection is not recoverable. This would imply the occurrence of at least one of the following:
\begin{itemize}
    \item None of the collection's guarantors are available or responsive. 
    \item The message delivery to and from the collection's guarantors is compromised by an attack, e.g., Routing Attack. 
    \item There is no time-bound on the message delivery to and from the collection's guarantors.
    \item There is no time-bound on the processing clock of the collection's guarantors.
\end{itemize}

As following the Honest Lower-bound assumption, there exists at least one honest Collector Node among the guarantors, having none of the collection's guarantors responsive or available contradicts the Honest Behavior assumption of the lemma stating the necessity on the honest actors to maintain their availability and responsiveness. Having the message delivery concerning the guarantors under attack contradicts the Message Delivery assumption of this lemma on the guaranteed message delivery to and from the guarantors. Having no bound on the processing clock of guarantors or the message delivery to and from them contradicts the Network Model assumption of the lemma. Hence, we conclude that as long as there exists at least one honest guarantor for a guaranteed collection under the assumptions of Lemma \ref{lemma:safety_1}, the collection remains available, accessible, and recoverable.
\vspace{-10pt}\begin{flushright}$\square$\end{flushright}

\begin{theorem}[Collection Text Availability\label{thm:safety}]$~$\\
Given a system with the following properties:
\begin{itemize}
    \item \textbf{Network Model:} a partially synchronous network with message traverse time bound by $\Delta_{t}$ and the relative processing clock of the nodes bound by $\phi_{t}$. Each node is assumed to have a proper defence and protection against network attacks (e.g. Denial-of-service attacks). 
    \item  \textbf{Message Delivery:} a fault-tolerant message routing functionality that guarantees the message delivery with high probability 
    \item \textbf{Stake Distribution:} equally staked Collector Nodes 
    \item \textbf{Byzantine Fraction:} more than $\frac{2}{3}$ of Collector Nodes' stakes are controlled by honest actors
    \item \textbf{Honest Behavior:} honest actors follow the protocols as specified, and maintain their availability and responsiveness in the system over time
\end{itemize}
There exists a configuration of \Architecture where each guaranteed collection is available with a high probability. 
\end{theorem}

\noindent
\textbf{Proof of Theorem \ref{thm:safety}:}  \\
In our previous white paper \cite{hentschel2019flow}, we have shown that by having $\collectorsNum = 1040$, 1/3 of these nodes being Byzantine, and our typical cluster size of $[50, 80]$ nodes, the probability  of having a Byzantine cluster is very unlikely. 

Despite \Architecture's proactive architectural setup for maintaining the availability of the collections, \Architecture also introduces a reactive mechanism (i.e., \techterm{Missing Collection Challenge}\footnote{%
    The analysis of the Missing Collection Challenge in \cite{hentschel2019flow} (specifically Proof of Theorem 3 therein) is more extensive than the proof presented here. Here, we assume that an honest actor will diligently but unsuccessfully query all guarantors for the desired collection before raising a Missing Collection Challenge. The analysis in \cite{hentschel2019flow} also covers the case of a lazy or trolling requester, which might query a subset of the guarantors and raise a challenge if the queried subset does not respond. 
}) to confirm and slash the Byzantine actors attributed to a missing collection  \cite{hentschel2019flow}.

\begin{theorem}[Liveness of Collection Formation\label{thm:liveness}]$~$\\
Given a system with the following properties:
\begin{itemize}
    \item \textbf{Network Model:} a partially synchronous network with message traverse time-bound by $\Delta_{t}$ and the relative processing clock of the nodes bound by $\phi_{t}$
    \item  \textbf{Message Delivery:} a fault-tolerant message routing functionality that guarantees the message delivery with high probability 
    \item \textbf{Stake distribution:} equally staked Collectors Nodes
    \item \textbf{Byzantine fraction:} more than $\frac{2}{3}$ of Collector Nodes' stakes are controlled by honest actors
    \item \textbf{Honest Behavior:} the honest actors follow the protocols as specified, and maintain their availability and responsiveness in the system over time
\end{itemize}
the collection formation always progresses. 
\label{flow:thm_collection_liveness}
\end{theorem}

\noindent
\textbf{Proof of Theorem \ref{thm:liveness}:}
By contradiction, assume that there is an unknown point of time $t_{stop}$ for which the collection formation process as described in this section stops working for all points of time $t > t_{stop}$. By collection formation being stopped, we mean that there is absolutely no single collection formed by any of the clusters of Collector Nodes once $t_{stop}$ elapses. This implies at least one of the following situations applies to the system by elapsing $t_{stop}$, which results in none of the clusters being able to reach a consensus over closing their collection: 
\begin{itemize}
    \item Less than $\frac{2}{3}$ of the Collector Nodes of \textbf{every} cluster are honest. 
    \item The message delivery to a fraction of at least $\frac{2}{3}$ of Collector Nodes of each cluster is compromised by an attack, e.g., Routing Attack. 
    \item There is no time-bound on the message delivery to and from a subset of at least $\frac{2}{3}$ of the Collector Nodes in each cluster.
    \item There is no time-bound on the processing clock of at least $\frac{2}{3}$ of Collector Nodes of each cluster.
\end{itemize}

Following the Honest behavior and Stake distribution assumptions of the theorem, having less than $\frac{2}{3}$ honest Collector Nodes contradicts the Byzantine fraction assumption of the theorem. Likewise, compromising the message delivery to a fraction of at least $\frac{2}{3}$ of the Collector Nodes on each cluster contradicts the Message Delivery assumption of the theorem. Finally, having no bound on either the message delivery time to or the processing time of at least $\frac{2}{3}$ of Collector Nodes in each cluster contradicts the Network Model assumption of the theorem. Hence, we conclude that as long as the assumptions of the theorem hold, the collection formation progresses. 

It is worth mentioning that, as discussed earlier, we envision it is rare for a Byzantine cluster to exist. Such a cluster may aim at compromising the liveness of the system by ignoring all the submitted transactions to that cluster. Nevertheless, upon detecting an attack on liveness by a user agent, the agent may simply try to switch the corresponding cluster of its transaction by changing some metadata of its transaction, e.g., changing the transaction's \texttt{ReferenceBlockHash} (see Listing \ref{message:tx}). This minor change results in the transaction's cluster assignment changing and so, it can be routed to a (non-Byzantine) cluster (see Equation \eqref{flow:eq_collection_assignment}) and hence processed towards inclusion in a collection.

\newpage
\section{Block Formation}
\label{flow:sec_consensus}
\subsection{Overview}
\label{flow:consensus_subsec_overview}
 \techterm{Block formation} is a continuous process executed by consensus nodes to form new blocks. The block formation process serves multiple  purposes: 
 \begin{itemize}
    \item Including the newly submitted guaranteed collections and reaching agreement over the order of them. 
    \item Providing a measure of elapsed time by continuously publishing blocks and determining the length of an epoch. 
    \item Publishing result seals for previous blocks of the chain. A block's result is ready to be sealed, once it is finalized by the Consensus Nodes, executed by the Execution Nodes, and the execution result is approved by the Verification Nodes  (see \cite{hentschel2019flow} for more details).
    \item Publishing slashing challenges and respective adjudication results.
    \item Publishing protocol state updates, i.e., slashing, staking, and unstaking. Whenever a node's stake changes, the Consensus Nodes include this update in their next block. 
    \item Providing a \entropy (see Section \ref{flow:subse_drb} for more details). 
\end{itemize}
In an unlikely event of no new guaranteed collection, consensus nodes continue block formation with an empty set of collections. In other terms, block formation process never gets blocked.

The block-formation process utilizes the underlying round-base consensus protocol we described in Section \ref{flow:subsec_consensus}. In \Architecture, we distinguish between \texttt{ProtoBlock}s and (proper)  \texttt{Block}s, which are formally defined in section \ref{sec:BlockFormation:DataStructures} below. The BFT consensus algorithm generates and finalizes \texttt{ProtoBlock}s. The only difference between a \texttt{ProtoBlock} and a (proper) \texttt{Block} is that a \texttt{ProtoBlock} does not contain any \entropy. The \entropy in a \texttt{Block} is required for various processes in \Architecture, including deterministically generating pseudo-random numbers during transaction execution, assigning collectors to clusters, etc. Hence, \texttt{ProtoBlock}s cannot be processed by nodes other than consensus nodes due to the lack of \entropy.

After a \texttt{ProtoBlock} is generated by the BFT consensus protocol, the Distributed Random Beacon (DRB) nodes sign the \texttt{ProtoBlock}. Their threshold signature over the \texttt{ProtoBlock} is the \texttt{Block}'s \entropy. The \texttt{ProtoBlock} plus the DRB's signature together form the (proper) \texttt{Block}.  While the Random Beacon is run by a set of consensus nodes, generating the randomness is \emph{not} part of the consensus process. Instead, the \entropy is added in a subsequent step to generate a proper \texttt{Block}.

\subsection{Data Structures\label{sec:BlockFormation:DataStructures}}

\subsection*{\texttt{ProtoBlock}s}

The BFT consensus algorithm works on the level of \texttt{ProtoBlock}. In each round, a consensus node is selected as \techterm{primary}, whose main task is to collect signatures for the last \texttt{ProtoBlock} and propose the next \texttt{ProtoBlock}. Once a \texttt{ProtoBlock} is proposed, the Consensus Nodes vote for its inclusion in the finalized chain. The structure of a well-formed \texttt{ProtoBlock} is specified in Listing \ref{message:Protoblock}. We provide definitions for the \texttt{ProtoBlock}'s individual fields below.

\begin{figure}[!t]
\begin{minipage}{\linewidth}
\begin{lstlisting}[language=protobuf2,style=protobuf, caption={ \Architecture's proto block structure}, label={message:Protoblock}]
message ProtoBlock {
  bytes  previousBlockHash;
  uint64 height;
  repeated GuaranteedCollection guaranteedCollections;
  repeated BlockSeal blockSeals; 
  SlashingChallenges slashingChallenges; 
  ProtocolStateUpdates protocolStateUpdates; 
  }
\end{lstlisting}
\end{minipage}
\end{figure}

\begin{itemize}[nosep]
    \item \texttt{previousBlockHash} points to the hash of the immediate predecessor of this block on the chain. By pointing to the previous block, blocks form a blockchain. 

    \item \texttt{height} corresponds to the current index of the block in the chain. 
    Formally, the block chain forms a tree with the blocks as vertices. The height of a block is the depth of the respective vertex in the tree. 
    
    \item  \texttt{guaranteedCollections} is an ordered list of new guaranteed collections that are introduced by this block. As explained in Section \ref{flow:sec_collection}, a guaranteed collection includes the hash value of the transactions in the collection, but not the transactions themselves. The transaction texts are stored by the Collector Nodes, who guaranteed the collection, and provided by them upon request.

    \item \texttt{blockSeals} is a list of hash commitments to the verified execution results of the previous blocks. Sealing a block happens after the submission of a minimum effective vote of $\frac{2}{3}$ of the block's result approvals from the Verification Nodes to the Consensus Nodes, with no challenges are pending. Check out our previous paper \cite{hentschel2019flow} for more details on execution and verification.
 
    \item \texttt{slashingChallenges} is the list of new slashing challenges that have been submitted to the Consensus Nodes for adjudication. Any staked node can submit slashing challenges against another node in the \Architecture network. For example, Verification Nodes may submit slashing challenges against the computation result of the Execution Nodes. 
    
    \item  \texttt{protocolStateUpdates} contains a cryptograpic commitment to the updated protocol state of the system as a result of executing this block. In \Architecture, all the staked nodes follow recently finalized blocks to update their view of the system and progress with their roles. Moreover, publishing this information in the blocks also acts as a public bulletin board of the changes in the system state.

    \item \texttt{aggregatedConsensusSigs} contains the aggregated signatures of the Consensus Nodes that voted in favor of this block. 
\end{itemize}

\subsection*{\texttt{Block}s}

The structure of a well-formed \texttt{Block} is presented by Listing \ref{message:block}. It contains commitments to all necessary information for progressing with transaction computation and evolving the protocol state in a deterministic manner. 
However, from a data perspective, a \texttt{Block} is \emph{not} self-contained.
For example, while cryptographic hashes of the execution state are included, the state data itself is not. 
This implies that anyone can verify the integrity of any subset of the execution state 
using the hashes in the \texttt{Block} (and merkle proofs which must be provided alongside the actual data).

\begin{figure}[!t]
\begin{minipage}{\linewidth}
\begin{lstlisting}[language=protobuf2,style=protobuf, caption={ \Architecture's block structure}, label={message:block}]
message Block {
  ProtoBlock protoBlock;
  Signature sourceOfRandomness;
}
\end{lstlisting}
\end{minipage}
\end{figure}

Essentially, a \texttt{Block} is simply a wrapper around a \texttt{ProtoBlock}, which adds the \texttt{sourceOfRandomness} to the block. Formally, the \entropy is a reconstructed threshold signature from the DRB Nodes (see Section \ref{flow:subse_drb} for details.)

\subsection{Block formation process}
\Architecture requires that all Consensus Nodes follow the same protocol when responding to events. 
We describe the block formation process steps from the perspective of a Consensus Node which we denote by $\texttt{this}$. 
A Consensus Node can hold two separate secret keys: 

\begin{itemize}
    \item \texttt{$\texttt{sk}_{\texttt{this}}^{\texttt{stake}}$} is used for staking.
    \item If the Consensus Node is a member of the DRB, it has a second secret key \texttt{$\texttt{sk}_{\texttt{this}}^{\texttt{DRB}}$} which is used for generating the node's threshold signature shares.
\end{itemize}
For the node $\texttt{this}$ in the \Architecture network, its role, public staking key \texttt{$\texttt{pk}_{\texttt{this}}^{\texttt{stake}}$}, current stake amount, and the node's network IP address \texttt{$ip\_addr_{\texttt{this}}$} are known to all other nodes. Furthermore, if the node is a member of the DRB committee, also its public key \texttt{$\texttt{pk}_{\texttt{this}}^{\texttt{DRB}}$} is publically known. 

\subsubsection{Block Proposal}
Following \Architecture's underlying consensus protocol, Consensus Nodes are selected as \techterm{primaries} with a probability proportional to their stake using a probabilistic but verifiable protocol. 
The Consensus Node selected as the \techterm{primary} of the current round has to propose the next \texttt{ProtoBlock} $pb$. It includes the guaranteed collections, block seals, slashing challenges, protocol state updates, and a commitment to the \techterm{protocol state} after the updates. 
Subsequently, it broadcasts the proposed \texttt{ProtoBlock} $pb$ to all other Consensus Nodes. 
As a security measure, the underlying consensus protocol has a timer to detect the primary's failure. If the primary does not generate a block within this time, the Consensus Nodes progress to the next round and \techterm{primary}. 
 
As mentioned earlier, if there are no guaranteed collections available, the primary should produce a \texttt{ProtoBlock} $pb$ with an empty list of guaranteed collections (i.e., $b.\texttt{guaranteedCollections} = \emptyset$), but still include the potential block seals, protocol state updates, challenges, adjudications, etc. It is worth noting that the Consensus Nodes only work with and decide upon the order of the collection hashes but not the full collection texts. Hence, they do not need to inspect the collections' transactions unless an execution result is being challenged \cite{hentschel2019flow}.

\subsubsection{Block Proposal Evaluation and Voting}
After the \techterm{primary} broadcasts its \texttt{ProtoBlock}, the rest of Consensus Nodes evaluate the proposal and vote. Each Consensus Node follows the consensus steps for this round (see Section \ref{flow:subsec_consensus}), if it approves the proposed \texttt{ProtoBlock}. 
Each Consensus Node runs Protocol \ref{protocol:BlockProposalEvaluation} and only votes in favor of a \texttt{ProtoBlock} if all the listed conditions are satisfied. In Protocol \ref{protocol:BlockProposalEvaluation}, we denote the current Consensus Node as `\texttt{this}'.\\

\begin{protocol}{\Architecture Consensus Protocol\label{protocol:BlockProposalEvaluation}}
\textbf{Input}

 \quad A proposed \texttt{ProtoBlock} $pb$ 

\textbf{Procedure}

    \quad Check these conditions:\smallskip
    \begin{enumerate}[nosep]
        \item $pb$'s proposer  is the selected \texttt{primary} of this round and $pb$ is signed by this node.
        \item $pb$ extends the known blockchain and is a descendent of the genesis block (without missing blocks).
        \item $pb$ is safe to vote on according to the rules of the consensus protocol. 
        \item $pb.\texttt{guaranteedCollections}$ contains a set of new guaranteed collections or is empty.
        \item \texttt{this} has received all guaranteed collections in $pb.\texttt{guaranteedCollections}$. 
        \item All the guaranteed collections in $p b.\texttt{guaranteedCollections}$ are authenticated (as defined in section \ref{flow:subsec_collection}).
        \item \texttt{this} has received all block seals in $pb.\texttt{blockSeals}$. 
        \item All the block seals in $pb.\texttt{blockSeals}$ are valid according to the conditions in \cite{hentschel2019flow}.
        \item \texttt{this} has received and verified all slashing challenges in $pb.\texttt{slashingChallenges}$. 
        \item Apply the protocol state updates listed in $pb$ to the final protocol state of the parent block (referenced by $pb.$\texttt{previousBlockHash}). Verify that the resulting protocol state matches the commitment in the block $pb$.
    \end{enumerate}
    \smallskip
     \quad  If all of these conditions are satisfied, vote in favour of $pb$. 
\end{protocol}

\subsubsection[Finalizing a Proto Block]{Finalizing a \texttt{ProtoBlock}}
Consensus Nodes sign a \texttt{ProtoBlock} to indicate their approval. 
In the current architecture of \Architecture, we implement \consensus, which is a BFT consensus algorithm with deterministic finality. However, without loss of generality, any other consensus algorithm that is BFT and has deterministic finality is suitable. In HotStuff, a \texttt{ProtoBlock} is finalized when there are three \texttt{ProtoBlock}s built on top of it. In order to build on top of a \texttt{ProtoBlock}, HotStuff requires more than $\frac{2}{3}$ effective votes from Consensus Nodes for it.  The safety of the HotStuff protocol guarantees that never two or more competing blocks are finalized.

\subsubsection{Source of randomness Attachment}
In \Architecture, a reliable and verifiable  randomness is essential for the system's Byzantine fault tolerance. The \texttt{sourceOfRandomness} field of a \texttt{Block} is used by \Architecture nodes to seed multiple pseudo-random-number generators. 
To provide a reliable source of randomness, \Architecture's DRB closely follows Dfinity's proposal \cite{hanke2018dfinity}. Protocol \ref{protocol:DistributedRandomBeaconSteps} specifies the steps for generating the \entropy for a \texttt{ProtoBlock}. 
Participating Consensus Nodes follow a two-step procedure to compute \texttt{sourceOfRandomness} for a given block. First, they sign the hash of the \texttt{ProtoBlock} and share their signatures with the other DRB nodes.  Second, a DRB node waits for the signature shares from other DRB members. Upon having more than $t$ distinct threshold signature shares over the hash of the \texttt{ProtoBlock}, each DRB node can recover the full threshold signatures over the hash for the \texttt{ProtoBlock}. The threshold signature is the \entropy of the block and its digest is used to seed the pseudo-random generators. Once the \entropy is ready, a DRB node broadcasts the (proper) \texttt{Block} to the entire network.   In Protocol \ref{protocol:DistributedRandomBeaconSteps}, we denote the current Consensus Node as `\texttt{this}'.\\

\begin{protocol}{\Architecture Distributed Random Beacon steps\label{protocol:DistributedRandomBeaconSteps}}
\textbf{Input}

 \quad A \texttt{ProtoBlock} $pb$

\textbf{Procedure}
    \begin{enumerate}
          \item \textbf{Step 1: Sign the proto block}
  \begin{enumerate}
    \item 
    $\sigma_{\texttt{this}} \leftarrow \texttt{Sign}(\texttt{Hash}(pb),\texttt{sk}_{\texttt{this}}^{\texttt{DRB}})$ 
    \item 
    Broadcast $\sigma_{\texttt{this}}$ to all DRB nodes. 
  \end{enumerate}
  
  \item \textbf{Step 2: Wait for $(t+1)$-many distinct threshold signature shares} $\sigma_{i_1}, \sigma_{i_2}, ..., \sigma_{i_{t+1}}$%
  \begin{enumerate}
    \item 
    $\sigma  \leftarrow \texttt{Recover}(P_{i_1}, P_{i_2}, ..., P_{i_{t+1}},\hspace{3pt} \sigma_{i_1}, \sigma_{i_2}, ..., \sigma_{i_{t+1}})$ (see Section \ref{flow:subse_drb})
    \item create \texttt{Block} $b$ with $b.\texttt{sourceOfRandomness} = \sigma$ and $b.\texttt{protoBlock} = pb$
    \item broadcast $b$ to the entire network  
  \end{enumerate}
\end{enumerate} 

\end{protocol}

\subsection{Protocol State Updates}
For each block, one may view the \techterm{protocol state} as a key-value dictionary. A key in the protocol state corresponds to the public staking key $\texttt{pk}^{\texttt{stake}}$ of a node. The corresponding dictionary value stores all of the node's properties. Relevant properties for a staked node include the node's role, its staking amount, and its DRB key (if applicable).
The way \Architecture maintains its \techterm{protocol state} is akin to the way Ethereum handles its state: 
\begin{itemize}[nosep]
    \item The \techterm{protocol state} is directly maintained by the Consensus Nodes (which compare to miners in Ethereum).
    \item Each block contains a commitment to the resulting \techterm{protocol state} \emph{after} all protocol state updates in the block have been applied. 
\end{itemize}
\medskip

\noindent
Changes in the protocol state of a block propagate to the children of this block.   
Hence, the protocol state is a relative concept that is evaluated with respect to a base block. For example, having two consecutive blocks $A$ and $B$ where $B$ is the immediate child of $A$, one must evaluate the correctness of the protocol state updates as a transition from block $A$ to $B$. If there are forks in the unfinalized part of the chain, the protocol state might be different in the forks. 

The protocol state of a node may change due to slashing, staking, and unstaking. Any staked node can submit a slashing challenge against another node. The slashing challenges are adjudicated by the Consensus Nodes, and the protocol state of a node may change accordingly. For each epoch, there is a certain block height (i.e., protocol-determined parameter) before which all new staking requests for the next epoch must be submitted as transactions. We envision that in mature \Architecture this block height is set such that the staking period ends about one day (approximately $80\hspace{2pt}000$ blocks) before the new epoch. 

To stake, an actor submits a staking transaction which includes its public staking key.
Once the staking transactions are included in a block and executed by the Execution Nodes, a notification is embedded into the corresponding Execution Receipt. When sealing the execution result, the Consensus Nodes will update the protocol state of the staking nodes accordingly. 

For unstaking, a node submits a transaction signed by its staking key. Once an unstaking transaction is included in a block during an epoch, it discharges the associated node's protocol state as of the following epoch. The discharged stake of an unstaked node is effectively maintained on hold, i.e., it can be slashed but it is not returned to the unstaked node's account. The stake is returned to the unstaked node after a waiting period of at least one epoch. 
The reason for doing so is two-fold. First, detecting and adjudicating protocol violations might require some time. Hence, some delay is required to ensure that there is enough time to slash a misbehaving node before its stake is refunded. Second, to prevent a long-range attack wherein a node unstakes, and then retroactively misbehaves, e.g., a Consensus Node signing an alternative blockchain to fork the protocol.


\subsection{Adjudicating Slashing Challenges}
Consensus Nodes adjudicate slashing challenges and update the protocol state accordingly. On receiving a slashing challenge from a node (the `challenger') against a challenged node, the Consensus Nodes include it into a block. This is done to embed an official record of the slashing challenges in the blockchain. In case the challenge contains a full proof, the Consensus Nodes adjudicate it right away. Otherwise, they wait for the challenged node to reply accordingly. If the challenged node does not reply within a timeout, it is slashed right away. Otherwise, if the challenged party responds with data required to adjudicate the challenge, the Consensus Nodes process the data and slash whomever (i.e., challenger or challenged node) that is at fault. In both cases, the Consensus Nodes publish the adjudication result in the next block together with the protocol state updates. The stakes of the nodes are affected as the result of the updates.

\subsection{Correctness Proof}

HotStuff \cite{HotStuff:2018, HotStuff:2019:ACM} is at the heart of \Architecture's Consensus Protocol \ref{protocol:BlockProposalEvaluation}. Flow only imposes some additional requirements on the validity of blocks, which the HotStuff protocol is agnostic to. Therefore, the safety of \Architecture's consensus protocol follows from HotStuff's safety. We formalize this insight in the following Corollary \ref{corr:FlowConsensusSafety}.  

\begin{corollary}[Safety]\label{corr:FlowConsensusSafety}
Given a system with:
\begin{itemize}
    \item \textbf{Network Model:} partially synchronous network with message traverse time-bound by $\Delta_{t}$ and the relative processing clock of the nodes bound by $\phi_{t}$
    \item  \textbf{Message Delivery:} a fault-tolerant message routing functionality that guarantees the message delivery with probability close to 1.
    \item \textbf{Byzantine Fraction:} more than $\frac{2}{3}$ of Consensus Nodes' stakes are controlled by honest actors
    \item \textbf{Honest Behavior:} the honest actors follow the protocols as specified, and maintain their availability and responsiveness in the system over time
    \item \textbf{Consensus Protocol:} the Consensus Nodes follow HotStuff, which is a BFT consensus protocol with deterministic finality
\end{itemize}
Under these prerequisites, Protocol \ref{protocol:BlockProposalEvaluation} is safe against arbitrary behaviour of Byzantine nodes. Specifically, Byzantine actors cannot introduce errors by deliberately publishing or voting for faulty blocks. Furthermore, the consensus Protocol \ref{protocol:BlockProposalEvaluation} has deterministic finality.  
\end{corollary}

\bigskip\noindent
HotStuff \cite{HotStuff:2018, HotStuff:2019:ACM} is a general-purpose BFT consensus protocol which is agnostic to the payload contained in the block.  In other words, the detailed structure of the block payload as well as the corresponding validity requirements do not affect liveness of HotStuff as long as an honest Primary has the ability to always propose a block. In Flow's Consensus Protocol \ref{protocol:BlockProposalEvaluation}, there are no hard requirements on the amount of content a Primary has to put in the block. An honest Primary will only include valid content, or propose an empty block. Therefore,  liveness of Flow's Consensus Protocol \ref{protocol:BlockProposalEvaluation} is also immediately follows from HotStuff, as stated in the following Corollary \ref{corr:FlowConsensusLiveness}.  

\begin{corollary}[Liveness]\label{corr:FlowConsensusLiveness}
Given a system with:
\begin{itemize}
    \item \textbf{Network Model:} partially synchronous network with message traverse time-bound by $\Delta_t$ and the relative processing clock of the nodes bound by $\phi_t$
    \item \textbf{Message Delivery:} a fault-tolerant message routing functionality that guarantees the message delivery with probability close to 1
    \item \textbf{Byzantine Fraction:} more than $\frac{2}{3}$ of Consensus Nodes' stakes are controlled by honest actors
    \item \textbf{Honest Behavior:} the honest actors follow the protocols as specified, and maintain their availability and responsiveness in the system over time
    \item \textbf{Consensus Protocol:} the Consensus Nodes only vote on \texttt{\upshape{ProtoBlock}s} if they are safe according to the rules of the HotStuff consensus protocol
\end{itemize}
Under these prerequisites, Protocol \ref{protocol:BlockProposalEvaluation} is live, i.e., the protocol will continue to finalize valid blocks.
\end{corollary}

\newpage
\section{Execution}
\label{flow:section_execution}
Execution Nodes follow the new blocks as they are published by the Consensus Nodes. A conservative\footnote{%
    The \Architecture protocol allows Execution Nodes to compute unfinalized blocks. This might allow the Execution Nodes to gain more revenue in an incentive model that specifically rewards speed of execution. However, such an optimistic Execution Node risks investing resources for computing blocks which are abandoned later. Furthermore, in contrast to finalized blocks, the BFT consensus algorithm does not guarantee the validity of unfinalized block proposals. Hence, an optimistic Execution Node must protect itself against processing invalid blocks. 
} Execution Node only processed finalized blocks to avoid wasting resources on unfinalized blocks, which might be abandoned later. While only Consensus Nodes actively participate in consensus, all other nodes still verify for themselves that a block is finalized according to the criteria of the BFT consensus protocol.

\begin{figure}[!b]
\centering
\includegraphics[scale=0.37]{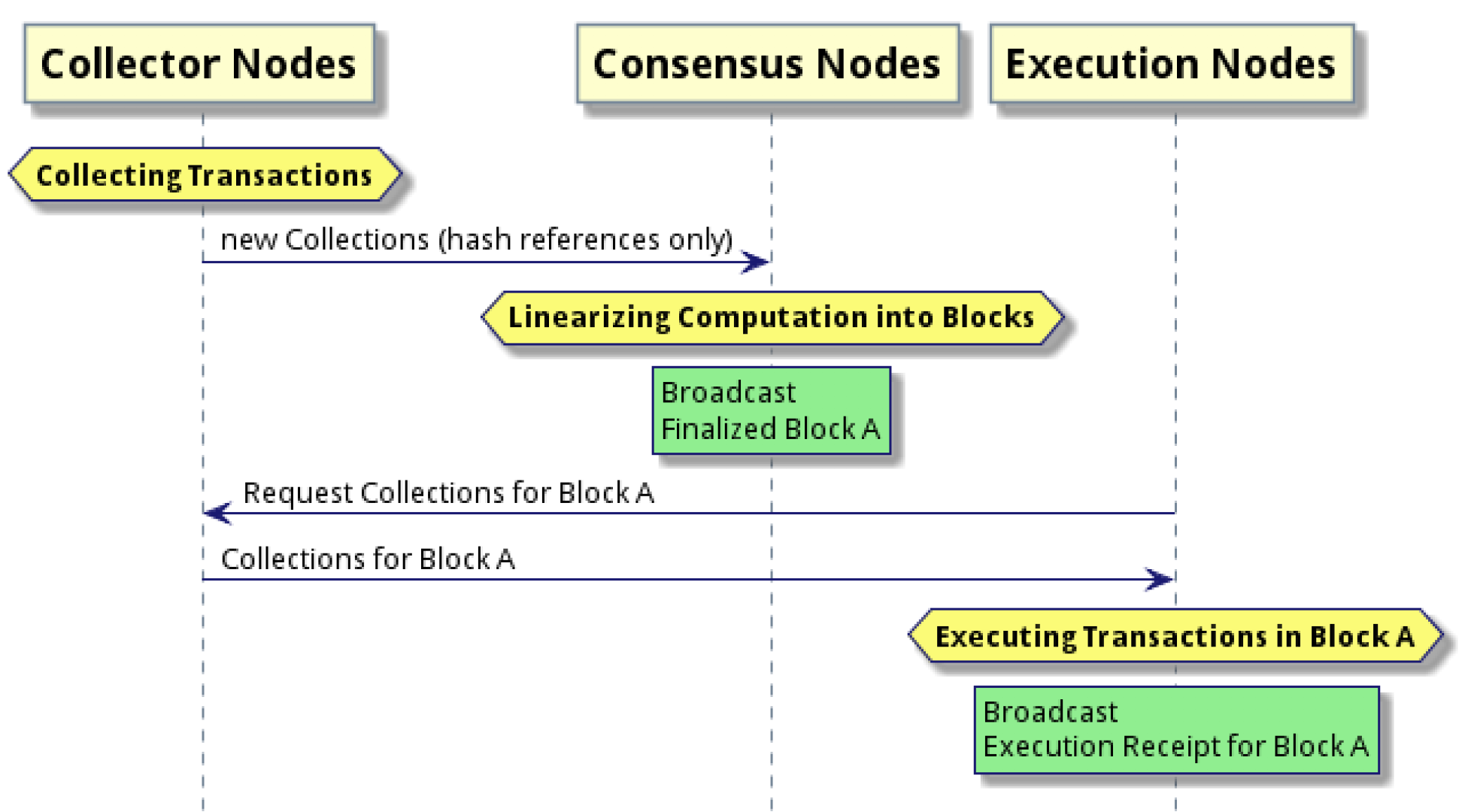}
\caption{Lifecycle of a transaction from collection until the execution. The yellow hexagons indicate the start of the individual stages in the \Architecture pipeline, e.g., the block sealing (i.e., Transactions Collection). The arrows show the inter-role message exchange between the nodes. Green rectangles correspond to the broadcasting events, in which a message is disseminated to all of the staked nodes. For the sake of simplicity, we represent the normal flow of the transaction lifecycle and ignored the message exchanges related to adjudicating slashing challenges.}
\centering
\label{flow:figure_execution_path}
\end{figure}

\subsection{Overview}
\label{flow:subsec_execution_overview}
As illustrated by Figure \ref{flow:figure_execution_path}, in \Architecture, the \emph{execution} process is triggered when the Consensus Nodes broadcast evidence that the next block is finalized. It completes when the Execution Nodes broadcast their Execution Receipts.
The Execution Nodes are responsible for receiving a finalized block, executing its transactions, and updating the execution state of the system. During this process each Execution Node takes the following steps: 
\begin{itemize}[itemsep=0pt]
    \item Retrieves each collection of transactions that appeared in the finalized block from the relevant clusters.
    \item Executes the transactions according to their canonical order in the finalized block and updates the execution state of the system. 
    \item Builds an \techterm{Execution Receipt} for the block, i.e., a cryptographic commitment on the execution result of the block
    \item Broadcasts the Execution Receipt to Verification and Consensus Nodes. 
\end{itemize}
We elaborate on each of these steps in the rest of this section. For the sake of simplicity, we describe the process from the perspective of a \emph{single} Execution Node. In the live system, the Execution Nodes operate completely independently from each other and execute the protocol in parallel.

\subsection{Collection Retrieval}
\label{flow:subsec_execution_collection_retrieval}
A guaranteed collection, as listed in the block (compare Listing \ref{message:collection}), only contains a hash of the set of transactions that it represents but not their full texts. As explained in Section \ref{flow:sec_collection}, the Collector Nodes that sign a guaranteed collection are also responsible sending its full text to other nodes upon request. 

Once an Execution Node confirms for itself that a block is finalized (see Section \ref{flow:sec_consensus}), it asks each cluster for their respective collection text so it can begin processing the transactions in order.
If the responsible guarantors do not provide a guaranteed collection, a node can issues a \techterm{Missing Collection Challenge} against the guarantors. The challenge is submitted directly to the Consensus Nodes. Missing Collection Challenges are discussed in detail in the third Flow Technical Paper  \cite{hentschel2019flow}.

When an Execution Node retrieves the transactions for a guaranteed collection, it 
reconstructs the collection's hash in order to verify that the transaction data is consistent with the guaranteed collection. Provided verification succeeds, a collection is considered \techterm{successfully recovered}.

\subsection{Block Execution}
\label{flow:subsec_execution_block_execution}
An Execution Node computes a finalized block $b$, when the following conditions are satisfied:
\begin{itemize}
    \item The execution result of the previous block to $b$ (the one referened by $b.\texttt{previousBlockHash}$) is available from either the Execution Node itself or from  another Execution Node. 
    \item All guaranteed collections in $b.\texttt{guaranteedCollections}$ have been successfully recovered. Alternatively, the Execution Node has a \techterm{Missing Collection Attestation}, which allows it to skip a lost collection (details are discussed in detail in the third Flow Technical Paper  \cite{hentschel2019flow}).
\end{itemize}


\noindent
Once the Execution Node has computed block $b$, it broadcasts an \techterm{Execution Receipt} to Verification and Consensus Nodes. Listing \ref{message:executionreceipt} shows the Execution Receipt's message format. In addition to the Execution Node's signature \texttt{executorSig}, an Execution Receipt has two types of data structures: an \techterm{Execution Result} (field \texttt{executionResult}), and a \techterm{Specialized Proof of Confidential Knowledge} (SPoCK) (field \texttt{Zs}). We detail each of these in the following sections \ref{sec:Execution:ExecResult} and \ref{sec:Execution:SPOCK}. 

\noindent
\begin{minipage}{\linewidth}
\begin{lstlisting}[language=protobuf2,style=protobuf, caption={Structure of an Execution Receipt in \Architecture. It consists of two primary data structures: an \texttt{ExecutionResult} and one or more \texttt{SPoCK}(s).}, label={message:executionreceipt}]
message ExecutionReceipt {
    ExecutionResult executionResult;
    repeated SPoCK Zs;
    bytes executorSig;
}
\end{lstlisting}
\end{minipage}

\subsubsection{Execution Result\label{sec:Execution:ExecResult}} 
An \techterm{Execution Result} for a block $b$ represents a commitment from an Execution Node to the interim and final states of computing the block $b$. As shown by Listing \ref{message:executionresult}, an Execution Result for a block contains the block hash (\texttt{blockHash}), hash reference to the previous Execution Result (\texttt{previousExecutionResultHash}), one or more chunks (\texttt{chunks}), and a commitment to the final execution state of the system after executing this block (\texttt{finalState}).

\begin{figure}[!h]
\centering
\includegraphics[scale=0.3]{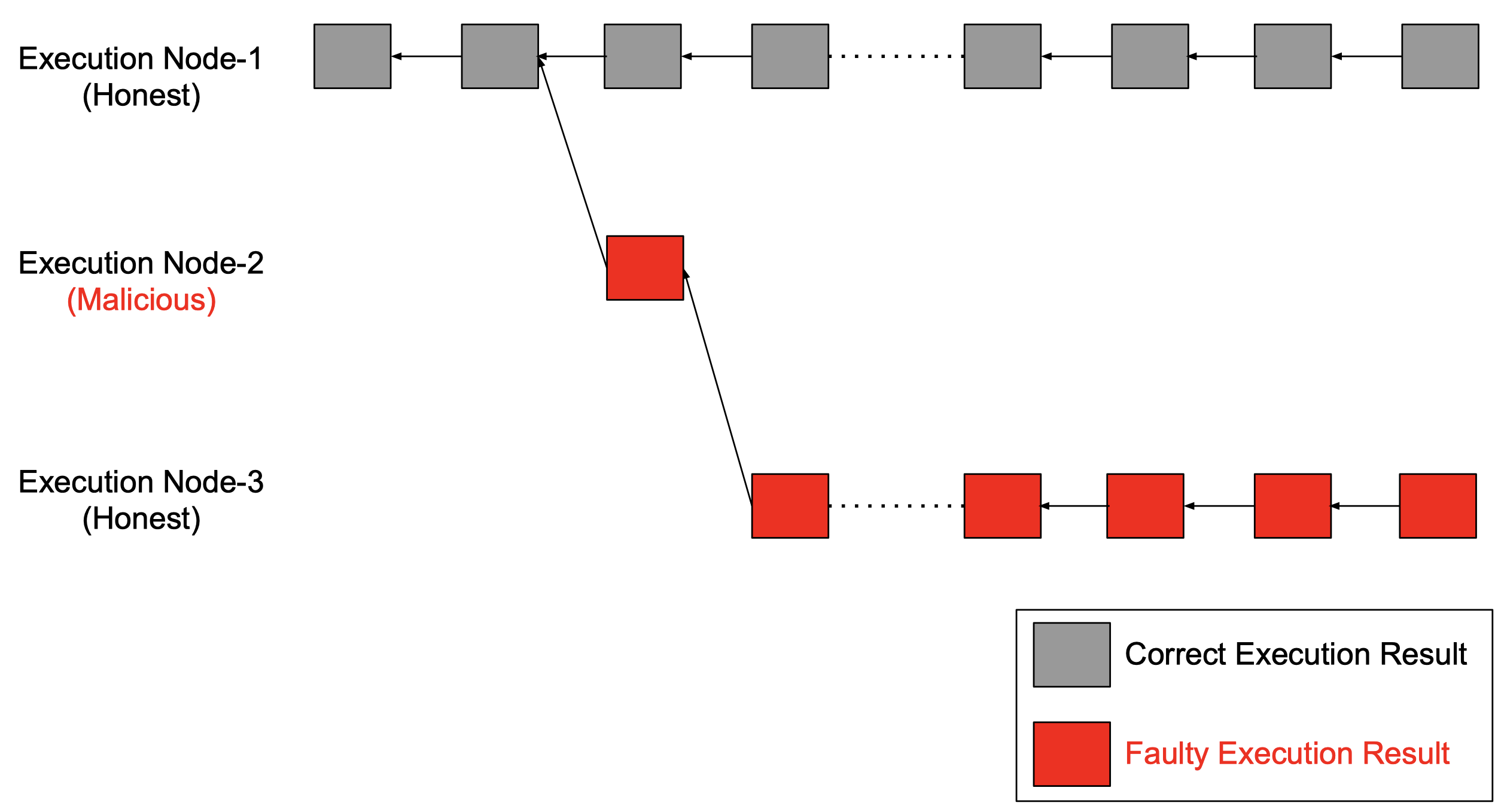}
\caption{The \texttt{previousExecutionResultHash} field of the Execution Results enables identifying and slashing the malicious Execution Nodes that originally injected an error into the system, instead of punishing the honest ones that only propagate the error. In this figure, the gray rectangles represent the correct Execution Results, and the red ones represent a faulty Execution Result. The arrows correspond to \texttt{previousExecutionResultHash} references, i.e., the Execution Result on the right side of an arrow takes its initial execution state from the result of the left one. The faults in the Execution Results that are generated by the honest Execution Node-3 are attributable to the malicious Execution Node-2 that initially injected them into its own Execution Result. In other words, \texttt{previousExecutionResultHash} of the Execution Results provides a chain of attributability that enables the Verification Nodes to trace back the errors, and identify and slash the malicious Execution Node-2 that originally introduced the error instead of punishing honest Execution Node-3 that only propagated the error.}
\centering
\label{flow:figure_previouser}
\end{figure}

The \texttt{previousExecutionResultHash} acts as a commitment to the starting execution state for executing the block. It allows tracing an execution error back to the actor that originally introduced it. Figure \ref{flow:figure_previouser} shows an example of this case. 
It is worth noting that this linked sequence of Execution Results could be called \textit{"a chain"}, but we avoid the word in order to avoid confusion with the primary \textit{"block chain"} maintained by the Consensus Nodes. Also, as shown by Listing \ref{message:executionresult} each Execution Result has a hash reference to the block that it is associated with. However, we skip the inclusion of the block references in Figure \ref{flow:figure_previouser} for the sake of simplicity. Once a Verification Node attributes a computation fault to an Execution Node, it issues a \techterm{Faulty Computation Challenge (FCC)} against the faulty Execution Node. Verification of Execution Results and adjudicating the FCCs are covered in detail in our other white paper \cite{hentschel2019flow}. 

Execution Results consist of one or more chunks. Chunking is the process for dividing a block's computation so such that the resulting chunks can be verified in a distributed and parallelized manner by many Verification Nodes. The execution of the transactions of a block is broken into a set of chucks. The chunking is done based on the computation consumption of the transactions, and in a way that the overall computation consumption of the transactions of chunks be similar to each other and do not exceed the system-wide threshold of $\Gamma_\textrm{chunk}$. Homogeneity of the chunks computation consumption is a security countermeasure against the injected faults in the computation-heavy chunks. In other words, if the computation consumption of the chunks is heterogeneous, the Verifier Nodes of the computationally heavier chunks may not be able to finish the verification before the sealing of the actual block associated with those chunks \cite{hentschel2019flow}. This enables the malicious Execution Nodes to attack the integrity of the system by injecting a computational fault into the computationally heavy chunks, which are likely to be left out of verification. 

The structure of a chunk is represented by Listing \ref{message:executionresult}, where \texttt{startStateCommitment} denotes a hash commitment to the execution state of the system \emph{before} the execution of the chunk. \texttt{startingTransactionCC} and \texttt{startingTransactionIndex} are the computation consumption and the index of the chunk's first transaction. \texttt{computationConsumption} denotes the overall computation consumption of the chunk. 
The chunking convention is: $\texttt{computationConsumption} \leq \Gamma_\textrm{chunk}$.

\noindent
\begin{minipage}{\linewidth}
\begin{lstlisting}[language=protobuf2,style=protobuf, caption={Structure of an Execution Result and Chunk in \Architecture}, label={message:executionresult}]
message ExecutionResult {
    bytes blockHash;
    bytes previousExecutionResultHash;
    repeated Chunk chunks; 
    StateCommitment  finalState;
}
message Chunk {
    StateCommitment  startStateCommitment;
    float   startingTransactionCC;
    uint32  startingTransactionIndex;
    float  computationConsumption;
}
\end{lstlisting}
\end{minipage}

\subsubsection{Specialized Proof of Confidential Knowledge (SPoCK)\label{sec:Execution:SPOCK}} 
The SPoCK is \Architecture's countermeasure against Execution Nodes copying Execution Results from each other instead of computing the block on their own. Also, SPoCKs are used to prevent Verification Nodes from blindly approving execution results without doing the verification work. SPoCKs are detailed in \cite{hentschel2019flow}, but it is sufficient for the purposes of this paper to understand that SPoCKs are cryptographic commitments, which are generated as part of the verification process to indicate that the Verification Node has done its job. Essentially, a SPoCK is a commitment to the execution trace for a single chunk. The field \texttt{ExecutionRecipt.Zs} is a list of SPoCKs, where \texttt{ExecutionRecipt.Zs[i]} is the SPoCK for the $i^{\textnormal{th}}$ chunk. It is worth emphasizing that \texttt{ExecutionRecipt.Zs[i]} can be generated by only executing the $i^{\textnormal{th}}$ chunk. 

\begin{figure}[!b]
\centering
\includegraphics[scale=0.3]{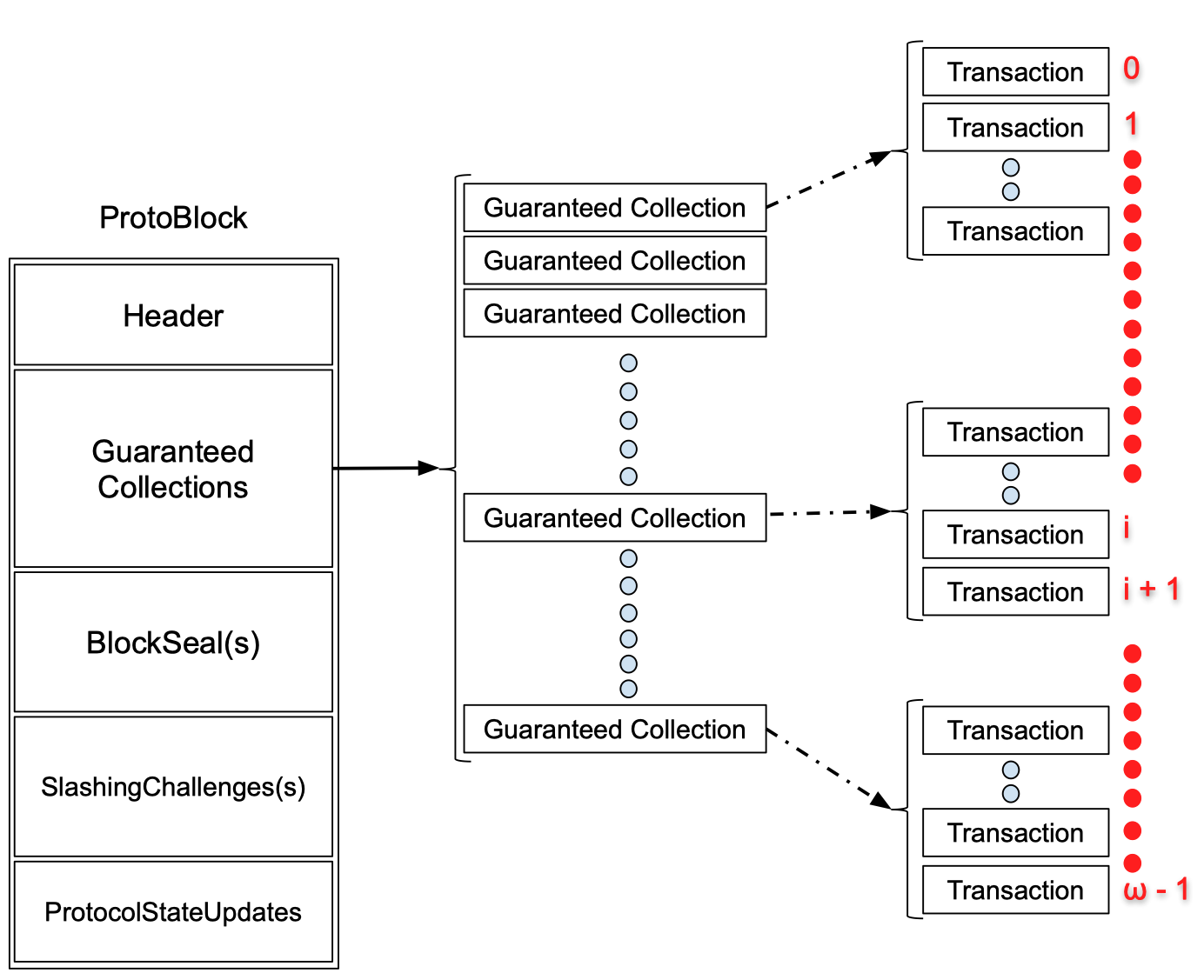}
\caption{The canonical order of the transactions in a \techterm{proto block} with $\omega$ transactions is shown in red. The transactions with the canonical orders of $1$ and $\omega$ are the first and last transactions in the sequential execution order of the \techterm{proto block}, respectively. The tick arrow corresponds to the \textit{inclusion} relationship, i.e., a \techterm{proto block} includes several guaranteed collections. The dashed arrows correspond to the referential relationship, i.e., each guaranteed collection holds a reference to a collection of several transactions.}
\centering
\label{flow:figure_canonical}
\end{figure}

\subsubsection{Execution Algorithm}
Algorithm \ref{flow:alg_block_execution} represents the \texttt{BlockExecution} algorithm that an Execution Node invokes individually for executing a block. The inputs to this algorithm are a finalized block $b$, which is ready for the execution, the hash of the previous Execution Receipt $h_{\texttt{er}_{prev}}$, and the execution state of the system before the execution of the block $b$, i.e., $\Lambda$. The output of the algorithm is an Execution Receipt for block $b$, which is denoted by $\texttt{er}_{b}$. The algorithm is invoked as follows: 
\begin{equation*}
    \texttt{er}_{b} =  \texttt{BlockExecution}(b, h_{\texttt{er}_{prev}}, \Lambda)
\end{equation*}
The Execution Node processes the transactions referenced by block $b$ in their \techterm{canonical order}. As shown by Figure \ref{flow:figure_canonical}, for the \techterm{proto block} associated with block $b$, the canonical order of the transactions is achieved by first sorting its guaranteed collections in the same order as they appear in the \techterm{proto block}, and then for each collection, sorting its transactions in the same order as the collection references. The first transaction of the first collection, and the last transaction of the last collection are then the first and last transactions of block $b$ based following the canonical ordering, respectively. To resolve the canonical order of the transactions associated with block $b$, the Execution Node invokes the function \texttt{canonical(b)}. The execution of each transaction $t_{i}$ is done by invoking the \texttt{execute} function, where $t_{i}$ is the $i^{\textnormal{th}}$ transaction of the block $b$ following its canonical order. On receiving the current execution state of the system $\Lambda$ and the transaction text $t_{i}$, the \texttt{execute} function returns the resulting execution state, the computation consumption of $t_{i}$, i.e., $\tau$, and an SPoCK for the execution of $t_{i}$, i.e., $\zeta$ (Algorithm \ref{flow:alg_block_execution}, Line \ref{flow:alg_block_execution_tx_execution}). \texttt{BlockExecution} keeps track of the computation consumption of the current chunk by aggregating the overall computation consumption of the executed transactions since the start of the current chunk (Algorithm \ref{flow:alg_block_execution}, Line \ref{flow:alg_block_execution_consumption}). 

For each transaction $t_{i}$, the Execution Node checks whether the current chunk's total computation consumption would exceed the threshold $\Gamma_\textrm{chunk}$ when transaction $t_{i}$ is added. If the threshold is not met, the transaction is considered as part of the current chunk, and the SPoCK of the current chunk (i.e., $\widetilde{\zeta}$) is updated by the SPoCK trace of $t_{i}$'s execution (Algorithm \ref{flow:alg_block_execution}, Line \ref{flow:alg_block_execution_update_trace}). 

Otherwise, if the threshold $\Gamma_\textrm{chunk}$ is reached for the inclusion of the transaction $t_{i}$ in the current chunk, the Execution Node closes the current chunk without including $t_{i}$. Accordingly, the transaction $t_{i}$ is taken as the first transaction of the next chunk. To exclude transactions without the need for reverting, the \texttt{BlockExecution} algorithm keeps track of the execution state by $\widetilde{\Lambda}$. Closing the current chunk is done by generating a commitment for the starting state of the chunk, which is tracked by $\Lambda_\textrm{start}$, and casting the chunk attributes into a \texttt{Chunk} message structure. 

The commitment for the start state is generated by Flow's Authenticated State Proof system, which provides three deterministic polynomial-time functions:
\begin{itemize}[nolistsep]
    \item \texttt{StateProofGen}
    \item \texttt{ValueProofGen}
    \item \texttt{ValueProofVrfy}
\end{itemize} 
\texttt{StateProofGen}$(\Lambda_\textrm{start})$ returns a verifiable hash commitment \texttt{startStateCommitment} to a given state snapshot $\Lambda_\textrm{start}$. Formally, Flow's execution state $\Lambda$ is composed of key-value pairs. The key $r$ can be thought of as the memory address and the value $v$ the memory content. 
As only a small fraction of state values (i.e., key-value pairs) are touched in a chunk, it would be inefficient to transmit the full state to the Verification Node. Instead, Execution Nodes transmit only the key-value pairs that are needed for checking their computation. In addition, an Execution Node also provides a proof to allow the Verification Node to check that the received key-value pairs are consistent with the state commitment in the chunk. The Execution Node generates these proofs by running $\gamma_x := \texttt{ValueProofGen}(\Lambda_\textrm{start}, r_{x})$. Here, $r_{x}$ denotes the key for an element of the state and $v_{x}$ the corresponding value. The Execution node sends the $(r_{x}, v_{x}, \gamma_x)$ to the Verification Node. The recipient can then verify that the values are indeed from the desired state by executing $\texttt{ValueProofVrfy}(r_{x}, v_{x}, \gamma_x, \texttt{startStateCommitment})$.
We leave the details of the Authenticated State Proof scheme as well as its construction to our future publications. Once the commitment to the starting state of the current chunk is ready, \texttt{BlockExecution} creates a new \texttt{Chunk} message (see Listing \ref{message:executionresult}).
The generated chunk is appended to the list $\mathcal{C}$, which holds the chunks of the block under execution. The SPoCK $\widetilde{\zeta}$ for the generated chunk is stored in the list $\texttt{Zs}$, which holds the SPoCKs for the chunks of the block. Once the current chunk is closed, the algorithm moves to the next chunk by re-initializing the variables that keep the attributes of the current chunk (Algorithm \ref{flow:alg_block_execution}, Lines \ref{flow:alg_block_execution_chunk_closing_start}-\ref{flow:alg_block_execution_chunk_closing_end}). $\tau_{0}$ keeps track of the computation consumption of the first transaction of the current chunk. 

Once all transactions of the block $b$ are executed in their canonical order and the corresponding chunks were created, \texttt{BlockExecution} creates an Execution Receipt ($\texttt{er}_{b}$) out of the hash of the block $b$ ($h_{b}$), the hash of the previous Execution Receipt ($h_{\texttt{er}_{prev}}$), the list of all the chunks of block $b$ ($\mathcal{C}$), the execution state of the system \emph{after} executing block $b$ ($\Lambda$), and the respective SPoCKs ($\texttt{Zs}$). The execution of block $b$ ends by each Execution Node individually broadcasting $\texttt{er}_{b}$ to the entire system.

\begin{algorithm}
{
\KwIn{\\
    \quad $b$: finalized block\\
    \quad $h_{\texttt{er}_{prev}}$: hash of the previous Execution Receipt\\
    \quad $\Lambda$: execution state of system prior to execution of $b$\\
    }
\KwOut{\\
        \quad $\texttt{er}_{b}$: Execution Receipt of block $b$\\
      }
    
\BlankLine
    \tcp{initializing the SPoCK list of the Execution Receipt}
    $\texttt{Zs} := [\,]$\;
    \tcp{initializing the list of chunks for block $b$}
    $\mathcal{C} := [\,]$\;
    \tcp{initializing start state of the current chunk}
    $\Lambda_\textrm{start} := \Lambda$\;
    \tcp{initializing starting transaction index of current chunk}
    $\texttt{startTransactionIndex} := 0$\;
    \tcp{initializing computation consumption of the current chunk}
    $c := 0$\; 
    \tcp{initializing the SPoCK of the current chunk}
    $\widetilde{\zeta} := \emptyset$\;
    \For{$t_{i} \in \texttt{\upshape canonical}(b)$}
    {
        \tcp{caching the state prior to execution of $t_{i}$ for chunk bookkeeping}
      $\widetilde{\Lambda} := \Lambda$\;
      $\Lambda, \tau, \zeta :=  \texttt{execute}(\Lambda, t_{i})$\;
      \label{flow:alg_block_execution_tx_execution}
      \If{$i = 0$}
      {
            $\tau_{0} := \tau$
      }
      \If{$c +  \tau > \Gamma_\textrm{\upshape{chunk}}$}
      {
            \label{flow:alg_block_execution_chunk_closing_start}
            \tcp{closing the current chunk}
            $\texttt{startStateCommitment} := \texttt{StateProofGen}(\Lambda_\textrm{start})$\;
            $chunk :=$ \textbf{new} \texttt{Chunk}$(\texttt{startStateCommitment},\, \tau_0,\, \texttt{startTransactionIndex},\, c)$\;
            $\mathcal{C}.\texttt{append}(chunk)$\;
            $\texttt{Zs}.\texttt{append}(\widetilde{\zeta})$\;
            
            \tcp{initializing the variables for the next chunk}
            $\Lambda_\textrm{start} := \widetilde{\Lambda}$\;
            $\texttt{startTransactionIndex} := i$\;
            $\tau_0 := \tau$\;
            $\widetilde{\zeta} := \emptyset$\;
            $c := 0$\;
            \label{flow:alg_block_execution_chunk_closing_end}
      }
      $c := c +  \tau$\;
      \label{flow:alg_block_execution_consumption}  
      \tcp{Updating the execution trace of the current chunk}
      $\widetilde{\zeta} := \texttt{TraceUpdate}(\widetilde{\zeta}, \zeta)$\;
      \label{flow:alg_block_execution_update_trace}  
    }
    
    \tcp{closing the last chunk}
    $\texttt{startStateCommitment} := \texttt{StateProofGen}(\Lambda_\textrm{start})$\;
    $chunk :=$ \textbf{new} \texttt{Chunk}$(\texttt{startStateCommitment},\, \tau_0,\, \texttt{startTransactionIndex},\, c)$\;
    $\mathcal{C}.\texttt{append}(chunk)$\;
    $\texttt{Zs}.\texttt{append}(\widetilde{\zeta})$\;
    $\texttt{er}_{b}$ := \textbf{new} \texttt{ExecutionReceipt}$(h_{b}, h_{\texttt{er}_{prev}}, \mathcal{C}, \Lambda, \texttt{Zs})$
    
\Indp
\Indm
\caption{BlockExecution}
\label{flow:alg_block_execution}
}
\end{algorithm}

\newpage
\subsection{Correctness Proof}
For completeness, we include the following safety Theorem \ref{Execution:SafetyTheorem}. For further details and the formal proof, the reader is referred to \cite{hentschel2019flow}.

\begin{theorem}[Safety]\label{Execution:SafetyTheorem}
Given a system with:
\begin{itemize}
    \item \textbf{Network Model:} a partially synchronous network with message traverse time-bound by $\Delta_{t}$ and the relative processing clock of the nodes bound by $\phi_{t}$
    \item  \textbf{Message Delivery:} a fault-tolerant message routing functionality that guarantees the message delivery with high probability 
    \item \textbf{Honest Behavior:}  honest actors which follow the protocols as specified and maintain their availability and responsiveness in the system over time
    \item \textbf{Byzantine Fraction:} at least one honest Execution Node
\end{itemize}
Even if all but one Execution Nodes are Byzantine and publishing the same faulty result, the faulty result is not sealed with a probability greater than $1 - \epsilon$ with $\epsilon \ll 1$. Depending on its system parameters, Flow can be tuned such that $\epsilon \leq 10^{-12}$. The system parameters are: the number of chunks in a block and the fraction of chunks of a block that each Verifier Node checks. 
\end{theorem}

\bigskip
\noindent
Assuming that there is at least one honest Execution Node in the system, a valid execution receipt will be published for each block. As HotStuff guarantees liveness of block formation, liveness of execution receipt generation is also guaranteed. We formalize this statement in the following Corollary \ref{Execution:LivenessTheorem}.

\begin{corollary}[Liveness]\label{Execution:LivenessTheorem}
Given a system with:
\begin{itemize}
    \item \textbf{Network Model:} a partially synchronous network with message traverse time-bound by $\Delta_{t}$ and the relative processing clock of the nodes bound by $\phi_{t}$
    \item  \textbf{Message Delivery:} a fault-tolerant message routing functionality that guarantees the message delivery with high probability
    \item \textbf{Honest Behavior:}  honest actors follow the protocols as specifiedand maintain their availability and responsiveness in the system over time
    \item \textbf{Byzantine Fraction:} at least one honest Execution Node 
\end{itemize}
The liveness block execution is guaranteed as the honest Execution Node(s) follow the specified algorithms and eventually, generate an Execution Receipt for the input block.  
\end{corollary}

\newpage
\section{Mitigating Attack Vectors}
\label{flow:section_attacks}
\subsection{Byzantine Clusters}
In mature \Architecture we envision clusters will be comprised of 20-80 randomly selected Collector nodes. 
While the probability for sampling a cluster with too many Byzantine nodes is small, it is unfortunately not negligible. 
A Byzantine cluster may attack the integrity of the system by injecting faulty transactions in their guaranteed collections.
It also may attack the liveness of the system by withholding the text of a guaranteed collection, to prevent the Execution Nodes from computing a block. 
The former attack on the integrity of the system is mitigated following the detectability and attributability design principle of \Architecture, i.e., the guarantors of a collection containing malformed transactions are slashed. The latter attack on the availability of collections is mitigated by introducing the Missing Collection Challenges (see \cite{hentschel2019flow} for details). This challenge is a slashing request against the Collector Nodes that originally guaranteed the availability of the missing collection. The challenge is directly submitted to the Consensus Nodes for adjudication. 

\subsection{Faulty Computation Result}
A malicious Execution Node may publish faulty \textit{Execution Receipt}s. However, we formally prove in \cite{hentschel2019flow} that faulty execution results are detected and challenged by the Verification Nodes with overwhelming probability. Upon detection of such faulty execution results, a Verification Node submits a slashing challenge against the faulty Execution Node to the Consensus Nodes (see \cite{hentschel2019flow} for more details). 


\clearpage
\addcontentsline{toc}{section}{Acknowledgments}
\section*{Acknowledgments}
Karim Helmy for his continued support and feedback on our technical papers.
\addcontentsline{toc}{section}{References}
\footnotesize
\bibliography{references}
\bibliographystyle{unsrt}

\end{document}